\documentclass[letterpaper,twocolumn,10pt]{article}
\usepackage{usenix2019_v3}

\usepackage{tikz}
\usepackage{amsmath}

\usepackage{booktabs}
\usepackage{graphicx}

\usepackage[labelfont=bf]{caption}
\usepackage{subcaption}
\usepackage{hyperref}
\usepackage{multirow}
\usepackage{bm}

\usepackage[inline]{enumitem}
\usepackage{amsthm}
\usepackage{array}
\usepackage{dirtytalk}
\usepackage[lined,boxruled]{algorithm2e}
\usepackage{bbm}
\usepackage{amssymb}
\usepackage[normalem]{ulem}
\usepackage{balance}
\usepackage{xcolor}
\usepackage[T1]{fontenc}
\usepackage{lipsum}

\usepackage{amsmath,amsfonts,bm}

\def\eqref#1{equation~\ref{#1}}

\def\1{\bm{1}}

\DeclareMathAlphabet{\mathsfit}{\encodingdefault}{\sfdefault}{m}{sl}
\SetMathAlphabet{\mathsfit}{bold}{\encodingdefault}{\sfdefault}{bx}{n}

\DeclareMathOperator*{\argmin}{arg\,min}

\newcommand{\myparagraph}[1]{
\vspace{0.08cm}
\noindent \textbf{#1}.
\hspace{0.1cm}
}

\usepackage{listings}
\newcommand{\modelname}[1]{{\fontfamily{lmtt}\selectfont #1}}

\newcommand{\data}{\mathcal{D}}
\newcommand{\bkg}{\mathcal{D}^\text{bkg}}
\newcommand{\real}{\mathbb{R}}
\newcommand{\expect}{\mathbb{E}}

\newcommand{\dataspace}{\mathcal{X}}
\newcommand{\labelspace}{\mathcal{Y}}
\renewcommand{\vec}[1]{\bm{#1}}
\newcommand{\vecw}{\bm{w}}
\newcommand{\vecx}{\bm{x}}
\newcommand{\vecy}{\bm{y}}

\newcommand{\spacex}{\mathcal{X}}
\newcommand{\spacey}{\mathcal{Y}}

\newcommand{\userset}{\mathbb{U}}
\newcommand{\clientset}{\mathbb{K}}

\newcommand{\anon}{\text{anon}}

\begin{document}

\date{}

\title{\Large \bf Understanding and Controlling Deanonymization in Federated Learning}

\author{
{\rm Tribhuvanesh Orekondy$^1$ \hspace{0.5cm} 
Seong Joon Oh$^{1\dagger}$ \hspace{0.5cm} 
Yang Zhang$^2$ \hspace{0.5cm} 
Bernt Schiele$^1$ \hspace{0.5cm} 
Mario Fritz$^2$} \vspace{0.5cm} \\
$^1$ Max Planck Institute for Informatics \hspace{0.25cm}
$^2$ CISPA Helmholtz Center for Information Security \\
Saarland Informatics Campus, Germany \\
} %

\maketitle
\footnotetext[2]{SJO is currently at Clova AI Research, Naver Corp.}

\begin{abstract}
Federated Learning (FL) systems are gaining popularity as a solution to training Machine Learning (ML) models from large-scale user data collected on personal devices (e.g., smartphones) without their raw data leaving the device.
At the core of 
FL is a network of anonymous user  devices sharing training information (model parameter updates) computed locally on personal data.
However, the type and degree to which user-specific information is encoded in the model updates  is poorly understood. 
In this paper, we identify model updates encode subtle variations in which users capture and generate data.
The variations provide a strong statistical signal, allowing an adversary to effectively deanonymize participating devices using a limited set of auxiliary data.
We analyze resulting deanonymization attacks on diverse tasks on real-world (anonymized) user-generated data across a range of closed- and open-world scenarios.
We study various strategies to mitigate the risks of deanonymization.
As random perturbation methods do not offer convincing operating points, we propose data-augmentation strategies which introduces adversarial biases in device data and thereby, offer substantial protection against deanonymization threats with little effect on utility.
\end{abstract}

\section{Introduction}

Advances in machine learning (ML) is increasingly fueled by accessibility to data sources capturing rich representations of the world e.g., 9M photographs \cite{openimages}, 1.6M tweets \cite{go2009twitter}, etc.
While such large-scale data advances learning fundamental  ML models (e.g., visual object recognition), the representations also encode a massive amount of unnecessary individual-specific information (e.g., person identities) \cite{orekondy2017towards,gurari2019vizwiz}.
For situations where the data is decentralized (e.g., user-generated photos on edge devices), Federated Learning \cite{McMahan2017CommunicationEfficientLO} provides a solution based on the principles of data minimization \cite{house2012consumer,gdprdataminim} 
towards training a ML model.
The core idea is participants distill from raw private data residing on individuals' device the information necessary to train the model, and intermittently communicate them to a server.
The information communicated by the participants take the form of model updates computed locally on-device.

\begin{figure}[t]
  \centering
  \includegraphics[width=\linewidth]{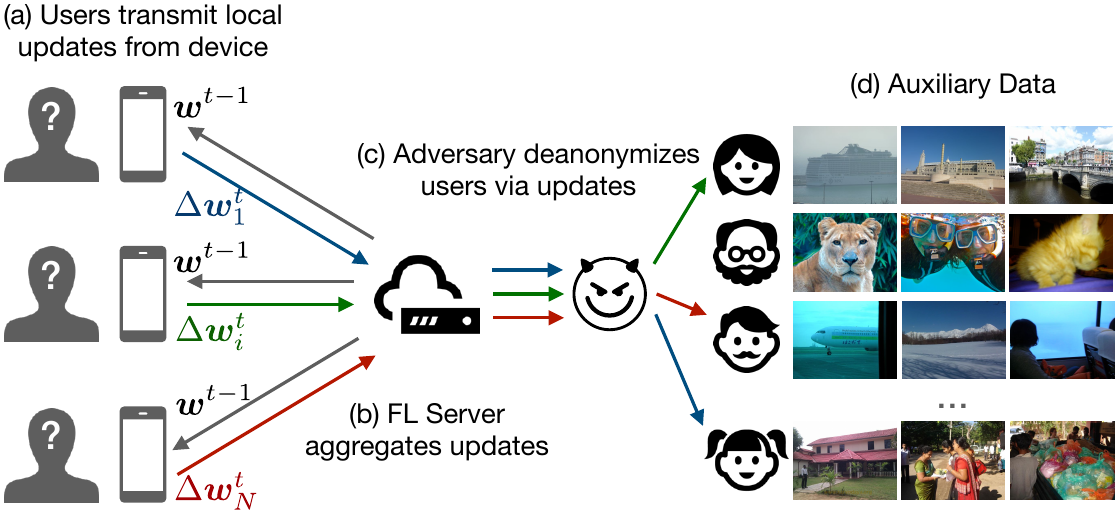}
  \caption{\textbf{Deanonymization in Federated Learning.} In this paper, we study how subtle user-biases captured in model parameter updates leads to deanonymization of their devices.}
  \label{fig:teaser}
\end{figure}

To prevent privacy violations, it is crucial that  model updates reveals information solely necessary for the training task (e.g., visual features to identify cats) and nothing about the participants (e.g., person identities).
To ensure this, federated learning is combined with additional steps to restrict the amount of data- and participant-specific information revealed in the process.
In the specific case of restricting participant-specific information encoded in model updates, typical steps include: stripping the data of PII information \cite{yang2018applied}, de-identifying the updates and auxiliary metadata \cite{yang2018applied,hard2018federated,McMahan2017CommunicationEfficientLO}, and avoiding authentication via user-identity prior to participation \cite{bonawitz2017practical}.
Hence, it is assumed that model updates received by the server contains minimal non-identifiable information to improve the model. %

However, it is in the nature of many real-world federated settings, that the clients represent diverse users with different interests, preferences and habits. Hence, the underlying data distributions of the users are not identically distributed and as a consequence, is characteristic of the users.
Therefore, we find that the model updates nonetheless encode individual-specific information and introduce \textit{significant} deanonymization risks.
Apart from constituting a privacy violation, deanonymization in federated learning undermines existing mechanisms to ensure the source of model updates are masked.
Furthermore, deanonymization amplifies effectiveness of recent inference attacks (e.g., attribute inference \cite{melis2018inference}), as identities can be tied to sensitive attributes inferred from the participants' private training data.

We investigate deanonymization risks and consequences by following the popular Federated Averaging algorithm \cite{McMahan2017CommunicationEfficientLO,bonawitz2019towards}, where participating devices intermittently communicate de-identified model parameter updates to a server. 
Here, the high-dimensional updates are a product of multiple gradient steps on multiple batches of the local device data.
We assume honest-but-curious server who intends to deanonymize participating devices (Fig.~\ref{fig:teaser}c) with limited access to prior information of users (Fig.~\ref{fig:teaser}d).
Central to our deanonymization attack is exploiting subtle, but inherent, individual- specific biases introduced when participants collect data on personal devices.
For instance, Alice capturing more photos of automobiles on her mobile device compared to Bob, who photographs food.
Our approach learns a suitable representation where the biases (modeled from limited prior data) can be leveraged to re-identify individuals via their model updates.

We evaluate deanonymization risks in a federated learning setup when training complex models (e.g., MobileNet CNNs \cite{howard2017mobilenets}) involving numerous participants (53-327 users).
 Furthermore, we use real-world (anonymized) user-generated datasets (e.g., PIPA, Blog) to closely emulate existing federated learning applications \cite{McMahan2017CommunicationEfficientLO,mcmahan2017learning}.
Our evaluation indicates that participants can be consistently deanonymized across a range of scenarios.
For instance, individuals transmitting model updates for an image classifier (with output classes e.g., chair, umbrella) on PIPA dataset are re-identified with high accuracy (19-175$\times$ chance-level).
Furthermore, we find the attacks surprisingly possible in spite of a range of data-limited scenarios, such as when the adversary has only a single prior example of the targeted individual.

Moreover, we propose a novel cross-modal attack which tackles a challenging scenario when the attacker's prior information varies in modality from the private data used during training by the participants.
For instance, the attacker leverages text information, while the participants are training using image data.
Our experiments indicate that in spite of the cross-modal challenge, attacks are quite effective (0.76 AUC).

It is worth noting that our deanonymization attack can also amplify the performance of recent attacks that infer sensitive properties of the training data.
For example, we show that learning an attack model to jointly perform deanonymization and attribute inference \cite{melis2018inference} are %
synergistic, with a consistent improvement of up to 4\% accuracy on both tasks. 
These results are further concerning, as sensitive attributes can be linked to identities of participants in federated learning.

After demonstrating the the risks of deanonymization in federated learning, we explore countermeasures to mitigate the threat.
We propose augmenting users' data distribution with an adversarial bias to decouple users' subtle variations from their prior information.
As a result, we propose the first mitigation strategy that directly operates on the user data itself, while maintaining utility of the task. %
We find our strategy mitigate attacks with up to 95\% effectiveness and incurs only negligible cost on the underlying task performance.
In contrast, we find perturbation- and DP-based training approaches (e.g., DP-FedAvg \cite{mcmahan2017learning}) incur large privacy and utility costs in our setup as they are typically effective only when training with a massive number of users (in the order of thousands).

\section{Related Work}
\label{sec:related_work}

In this section, we position our paper with existing literature on anonymization and privacy in ML.

\myparagraph{Deanonymizing (Insufficiently) Anonymized Data}
Organizations have largely believed that explicitly stripping away key identification information (e.g., names, SSN) from data records is sufficient to de-identify and provide anonymity of participating individuals.
Instances of this strategy to anonymize databases include Hospital Discharge dataset (GIC) \cite{sweeney1997weaving}, Netflix prize dataset \cite{bennett2007netflix}, and AOL search logs \cite{arrington2006aol}.
However, a long of line work, dating back to \cite{sweeney1997weaving,sweeney1997guaranteeing} highlight that although the de-identified database by itself might seem anonymous, joining with auxiliary publicly available data on a set of \textit{quasi-identifiers} (e.g., zip-code, gender) leads to effectively re-identifying individuals.
Consequently, in the aforementioned instances, many identities of participating individuals were re-identified using a public voter database \cite{sweeney1997weaving}, IMDb movie ratings \cite{narayanan2006break}, and search keywords \cite{barbaro2006aol} respectively.
This has motivated significant research in the area of identifying factors that potentially lead to deanonymization of individuals, such as in social networks \cite{narayanan2009anonymizing}, programmatic code \cite{abuhamad2018large}, and product reviews \cite{hernandez2018fraud}.
Research has also identified various sources of quasi-identifiers in unstructured data: profile attributes \cite{perito2011unique}, geo-location \cite{pyrgelis2017knock}, social graph structure \cite{korula2014efficient}, content \cite{goga2013exploiting}, stylometric features \cite{almishari2012exploring}, or RNA expressions \cite{backes2016privacy}.
In this paper, we tackle deanonymization of devices within Federated Learning, which enables users to anonymously participate towards the learning of an ML model using their private data.

\myparagraph{Attacks against Machine Learning Models}
Advances in ML has led to state-of-the-art statistical models being deployed `in the wild' to perform a variety of tasks such as autonomous driving, fraud detection, and medical diagnosis.
Attacks against such ML models can be targeted towards compromising the \textit{integrity} of the model (such as by evasion attacks 
\cite{biggio2013evasion,szegedy2013intriguing,CW17,ZE19,PMSW18,ACW18,GMXSX18,WYVZZ18,YVCZZ17}), or its \textit{privacy and confidentiality}.
Our focus is on the latter, since ML models need to obviously learn something as a result of training on (potentially private and confidential) data sources.
Attacks that compromise privacy of models in this setting include: model stealing \cite{lowd2005adversarial,tramer2016stealing,orekondy19knockoff}, membership inference \cite{shokri2017membership,salem2018ml,pyrgelis2017knock} which identifies if a particular example was used during training, attribute inference \cite{melis2018inference,ganju2018property} to identify properties that holds true for subsets of data, and model inversion \cite{fredrikson2015model,hitaj2017deep} to reconstruct training class exemplars.
In this work, we address a problem similar to membership and attribute inference, where we wish to identify properties that holds for subsets of data.
While membership inference intends to identify whether a particular \textit{example} was used during training, our adversarial goal can be cast as `userbase inference': to identify which particular \textit{individual} participated in training.%

\myparagraph{Attacks in Federated Learning}
Distributed ML on decentralized private user-generated data sources -- also referred to as Collaborative ML \cite{shokri2015privacy}, or Federated Learning \cite{McMahan2017CommunicationEfficientLO,bonawitz2019towards} -- is gaining popularity as it securely enables large-scale ML on private data sources.
While such approaches minimizes the privacy risks by keeping user in control of the raw private data, understanding the extent of privacy risks is gaining traction in the research community.
Understanding these risks in this setting is crucial since FL is designed to learn from private data spanning hundreds to tens of thousands of users.
Unfortunately, research in this area is minimal and work has only recently started to quantify these risks.

Given the considerable complexity of FL systems exposing many attack surfaces, we specifically focus on the anonymous \textit{gradient information} communicated by devices to the server.
One line of work studies malicious devices who exploit anonymity and secure aggregation protocols to mount poisoning and backdoor attacks \cite{bagdasaryan2018backdoor,bhagoji2019analyzing,xie2019dba} on the system.
Orthogonal to this line of work, are attacks \cite{nasr2018comprehensive,melis2018inference} where the server is modeled as an adversary instead of the device.
Since the server requires raw unencrypted access to  gradient signals for aggregation, it opens up threats to mount inference attacks that violate users' privacy.
Recently, \cite{nasr2018comprehensive} comprehensively explored membership inference on gradient parameters, including an analysis in an FL setting.
While our work explores a similar idea -- membership on a user-level -- we aim to determine it without access to the \textit{exact} training example(s) belonging to the user.
A closely related work to our paper is \cite{melis2018inference}, who propose an \textit{attribute inference attack} i.e., using the aggregated gradient signal to infer certain sensitive attributes (e.g., gender, race) that is not significantly correlated with the main task trained by participating users (e.g., sentiment analysis, gender classification).
In this work, we show that deanonymization complements and amplifies such attribute inferences, by enabling an adversary to additionally associate the sensitive attributes to an individual.

\section{Background, Notation and Terminology}
\label{sec:background}

In this section, we provide the preliminaries to Federated Learning, within which we explore our threat model in the next section.
At this point, we remark that research towards a Federated Learning system encompasses among many other things, architecture \cite{bonawitz2019towards}, optimization techniques \cite{konevcny2016optimization,McMahan2017CommunicationEfficientLO}, strategies to improve communication  \cite{konevcny2016communication}, aggregation \cite{bonawitz2017practical}, implementation \cite{abadi2016tensorflow}, and applications \cite{chen2019federated,yang2018applied,hard2018federated}.
To keep the background in this section concise, we present key concepts to understand:
(i) how devices generate model parameter updates using the \texttt{FederatedAveraging} \cite{mcmahan2017learning} algorithm; and 
(ii) how users anonymously communicate the parameter updates to the server in FL \cite{bonawitz2019towards,melis2018inference,nasr2018comprehensive}.

\myparagraph{Notation and Learning Objective}
In supervised learning, the overall objective is to learn a mapping $f_{\vecw}: \dataspace \rightarrow \labelspace$ of a model $f$ parameterized by $\vecw \in \real$.
The idea is to learn the parameters which minimizes the empirical risk represented by a loss function $L$
on a dataset $\data = \{ (\vecx_i, \vecy_i) \}_{i=1}^{n}$:
\begin{equation}
	\label{eq:central_loss}
	\hat{\vecw} = \argmin_{\vecw} H(\vecw) = \argmin_{\vecw} \frac{1}{n} \sum_i L(f_{\vecw}(\vecx_i),\ \vecy_i)
\end{equation}

In FL, data is partitioned across multiple devices $k \in \clientset$: $\data = \bigcup_k \data_k$.
Using $H_k(\vecw)$ to denote the objective solved locally on device $k$, the objective in Equation \ref{eq:central_loss} can now be re-written as:
\begin{equation}
	\label{eq:dist_loss}
	\hat{\vecw} = \argmin_{\vecw} \sum_{k=1}^{K} \frac{n_k}{n} H_k(\vecw)
\end{equation}

\myparagraph{Federated Averaging Algorithm}
Given the data $\data_k$ partitioned among devices $k \in \clientset$, the objective is to learn parameters $\vecw$ of the model $f_{\vecw}$, in the presence of a server $S$.
We use the popular \texttt{FederatedAveraging} algorithm  \cite{McMahan2017CommunicationEfficientLO,McMahan2017Collaborative} (Algorithm~\ref{algo:federated}) proposed specifically to perform training on non-IID and imbalanced decentralized data; this has also served as the footing for multiple prior works \cite{geyer2017differentially,mcmahan2017learning,bonawitz2017practical,smith2017federated}.
The idea here is that training occurs over multiple rounds, where in each round $t$, a fraction of devices $k \in \clientset_t$ train models $f_{\vecw}$ using the local data $\data_k^{\text{private}}$ and only communicate incremental model update $\Delta\vecw_k^t$ towards the server's global model $\vecw^t$.
The server aggregates (such as by averaging) parameter updates from multiple devices and shares back an updated improved model after each round.
Over multiple rounds of communications, the devices converge to model parameters $\vecw^T$ that has been effectively learnt from all the data $\data$, without their raw data ever being communicated to the server or another device.
It should be noted that although we consider the simple \texttt{FederatedAveraging} algorithm, we expect our results to generalize to a broad class of decentralized algorithms which involve periodically exchanging model parameter updates.

\begin{algorithm}[t]
\SetAlgoLined
\DontPrintSemicolon
\SetKwInOut{Input}{Input}
 \hspace{-0.5em} \textbf{Server's algorithm: }\;
   \KwIn{$K$ devices; $T$ number of rounds; $C$ fraction of devices sampled each round; $B$ device's batch size; $E$ number of local epochs}
   Randomly initialize $w^{t=0}$\;
   \For{round $t \leftarrow$ 1 \KwTo $T$}{
     $M \leftarrow \max(1,\; C \cdot K)$\;
     $\clientset_t \leftarrow $ sample $M$ devices from $\clientset$\;
     \For{$\text{client } k \in \clientset_t$}{
       $\Delta \vecw^{t+1}_k \leftarrow \texttt{DeviceUpdate}(k, \vecw^{t})$\;
     }
     $\vecw^{t+1} \leftarrow \vecw^{t} + \sum_{k \in \clientset_t} \frac{n_k}{n} \Delta \vecw^{t+1}_k$\;
   }
   
 \BlankLine
 \hspace{-0.5em} \texttt{DeviceUpdate}$(k, \vecw^t):$\;
 $\mathcal{B} \leftarrow $ split local data $\data_k^{\text{private}}$ into batches of size $B$\;
 $\vecw \leftarrow \vecw^t$\;
 \For{local epoch $i \leftarrow$ 1 \KwTo E}{
 	\For{batch $b \in \mathcal{B}$}{
    	$\vecw \leftarrow \vecw - \eta \nabla L(f_{\vecw}; b)$\;
    }
 }
 $\Delta \vecw \leftarrow \vecw^t - \vecw$\;
 \Return{$\Delta \vecw$}
 
 \caption{\texttt{FederatedAveraging} \cite{McMahan2017CommunicationEfficientLO} for training data on multiple devices}
 \label{algo:federated}
\end{algorithm}

\myparagraph{De-identification in FL}
A number of precautions are employed to ensure any identifiable information is stripped away from per-device update reports (which includes parameter updates $\Delta\vecw^t_k$ and additional metadata).
We first iterate over de-identification strategies employed on-device.
The client is initially registered into the FL process by being assigned population identifier \cite{yang2018applied} and thereby bypassing the need to authenticate with a device or user identity \cite{bonawitz2019towards}.
When possible, PII information is stripped away from the training data \cite{hard2018federated} prior to training on-device.
After a number of local training steps, the parameter updates $\Delta\vecw^t_k$ along with anonymized operational metrics \cite{yang2018applied} is transmitted by the device.
A (trusted) shuffler \cite{bittau2017prochlo} can be additionally employed to ensure the transmitted per-device update reports are further sanitized before reaching the server.
The shuffler typically strips away a range of user-specific metadata (e.g., IP addresses, routing details) and batches the reports (reordering updates to disassociate timing ordering information).
On the whole, multiple mechanism are in-place to ensure that only the essence of the update-reports (i.e., the parameter updates $\Delta\vecw^t_k$) are received by the server to aggregate updates.
Consequently, for the rest of the paper, we assume access to \textit{only} the parameter updates to perform deanonymization.

\section{Deanonymization Attacks in Federated Learning}
\label{sec:threat}
In this section, we begin by presenting our threat model to deanonymize devices. 
We then discuss an insight to why this threat arises and work towards our attack models.

\subsection{Threat Model}
\label{sec:threat_model}

To highlight deanonymization risks in Federated Learning \cite{McMahan2017CommunicationEfficientLO}, we analyze a scenario with $K$ honest users ($K \ge 2$) who collaboratively train an ML model $f_{\vecw}: \spacex \rightarrow \spacey$ over multiple rounds.
A server $S$ co-ordinates the training, by periodically collecting model updates from a random subset of users.
The model update communicated by each user is a result of performing multiple gradient steps over multiple batches on their local private data (see \texttt{DeviceUpdate(.)} in Algo. \ref{algo:federated}).
Furthermore, the model updates are stripped of identifiable metadata \cite{hard2018federated,bonawitz2017practical,yang2018applied} (e.g., device identifiers) and are optionally shuffled \cite{bittau2017prochlo} to obscure the source of each individual update.
Prior to summarizing information from multiple updates, we assume the server observes only the essence of per-user model update (i.e., parameter updates $\Delta \vecw_\text{anon}^t$) to improve $f_{\vecw}$.

We investigate deanonymization through the lens of an honest-but-curious server (the `adversary') during the training process who uses the model update as an attack surface.
The inference-time objective of the adversary is to deanonymize the model update i.e., re-identify the user $u$ who generated $\Delta \vecw^t_{\anon}$.
Such a deanonymization objective undermines sanitization mechanisms which de-identify model updates, such as decoupling the update from user identity \cite{bonawitz2019towards}, stripping away identifiable metadata \cite{hard2018federated,yang2018applied}, and blind-shuffling mechanisms \cite{bittau2017prochlo}.
Furthermore, deanonymization also serves as a stepping stone for amplifying information recovered from other inference attacks.
For instance, as we show later in \S \ref{sec:eval_att_inf}, deanonymization can be coupled with attribute inference attacks to improve attack performances and further associate recovered attributes with identities.

To deanonymize, the adversary leverages limited prior knowledge of users.
Formally, our threat model performs:
\begin{equation}
    \label{eq:threat1}
    f^{\text{adv}}: \Delta \vecw^t_{\anon} \times \data^{\text{prior}}_u \rightarrow u \stackrel{?}{=} \anon
\end{equation}
Here, $\Delta \vecw^t_{\anon}$ is the deanonymization target, which is a result of an anonymous user taking multiple gradient steps on her local data $\data^{\text{private}}_{\anon}$.
The adversary's auxiliary knowledge of users is denoted by  $\{\data^{\text{prior}}_u: u \in \userset\}$.
We assume $\data^{\text{prior}}_u$ represents a limited set of data generated by user $u$ and is distinct from their private data i.e., $\data^{\text{prior}} \cap \data^{\text{private}}_u = \emptyset \; \forall u \in \userset$.
For instance, historical data collected by the service, or content publicly shared by the users.
In Section \ref{sec:threat_adv_know}, we further elaborate on how we model the adversary's prior knowledge, as it plays a significant role in deanonymization attacks.

\subsection{Selection Bias and Biased Estimators}
\label{sec:threat_bias}
The core idea of our threat model is to use users' selection bias as an identification cue, which we hypothesize (and shortly verify) is consistent among both the users' prior data (known to adversary) and private device data (unknown to adversary).
This implicit user selection biases arise from behavioral factors \cite{fadem2012behavioral,hernan2004structural,berk1983introduction} that results in subtle variations of how humans capture data.
For instance, Alice's interest in automobiles might result in more variations of cars captured in her text/photos, compared to Bob whose interest lies in sports.
At this point, we remark that this results in a non-IID data distribution among data on users and devices, which is well-known in FL literature \cite{McMahan2017CommunicationEfficientLO,bonawitz2019towards}.
However, we do identify and exploit the property that although the data is  non-IID among users (large inter-user distances), the data displays lesser variation \textit{within} data generated by the same user (small intra-user distances).

\begin{figure}
  \centering
  \includegraphics[width=0.49\linewidth]{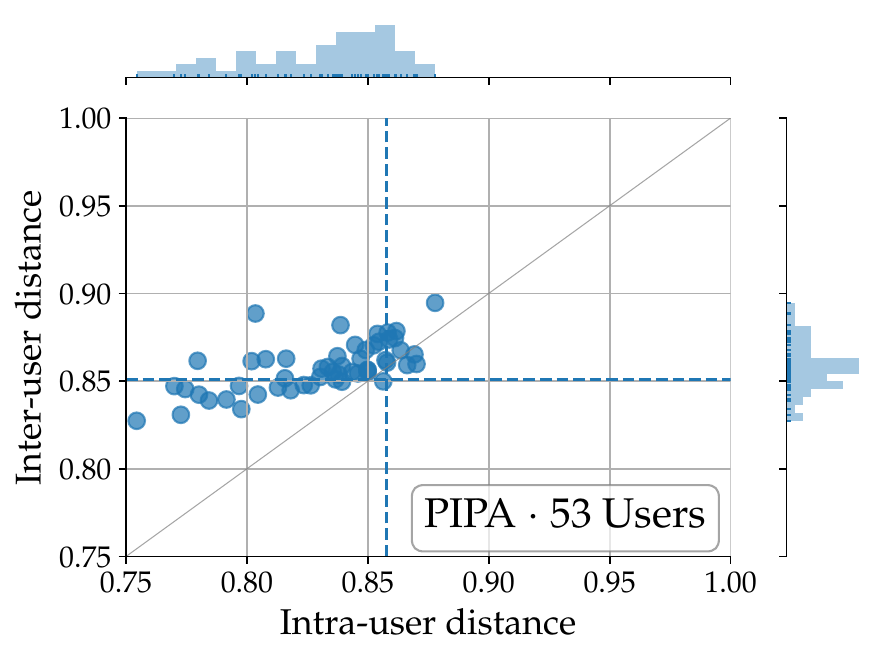}
  \includegraphics[width=0.49\linewidth]{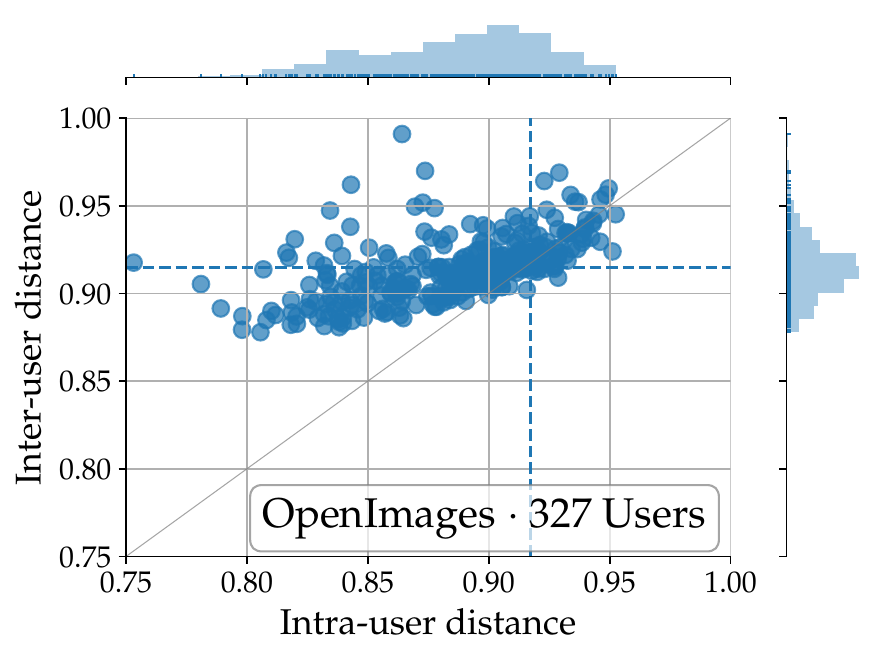}
  \caption{\textbf{Variations in user data.} Each point represents distances computed over the image set of a single user.
  }
  \label{fig:dataset_distances}
\end{figure}

\begin{figure*}[t]
  \centering
  \begin{subfigure}[t]{0.2\linewidth}
  	\centering
  	\includegraphics[width=0.5\linewidth]{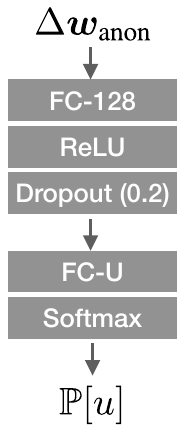}
  	\caption{MLP model for re-identification attack}
  	\label{fig:model_mlp}
  \end{subfigure}
  \hspace{2em}
    \begin{subfigure}[t]{0.2\linewidth}
  	\includegraphics[width=\textwidth]{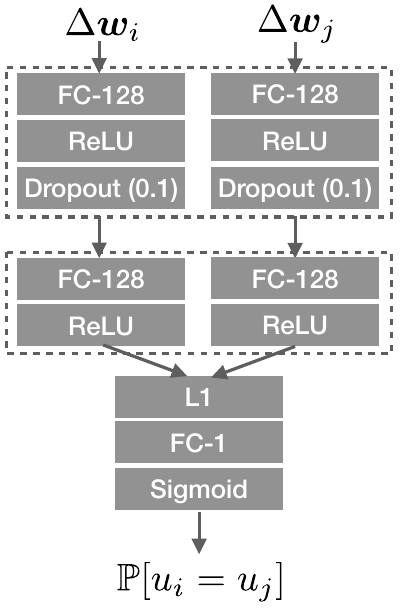}
  	\caption{Siamese model for matching attack}
  	\label{fig:model_siamese}
  \end{subfigure}
  \hspace{2em}
    \begin{subfigure}[t]{0.2\linewidth}
  	\includegraphics[width=\textwidth]{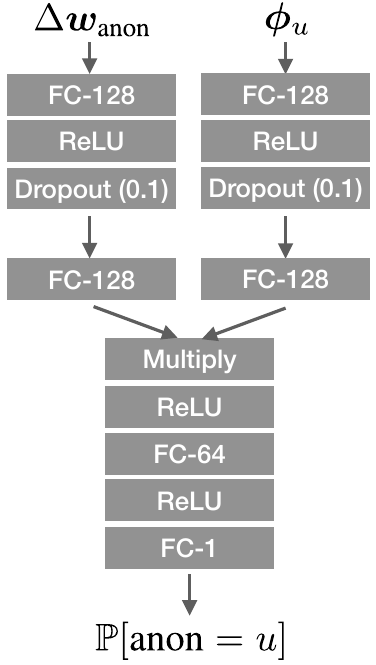}
  	\caption{Model for cross-modal matching attack}
  	\label{fig:model_crossmodal}
  \end{subfigure}
  \hspace{2em}
    \begin{subfigure}[t]{0.2\linewidth}
  	\includegraphics[width=\textwidth]{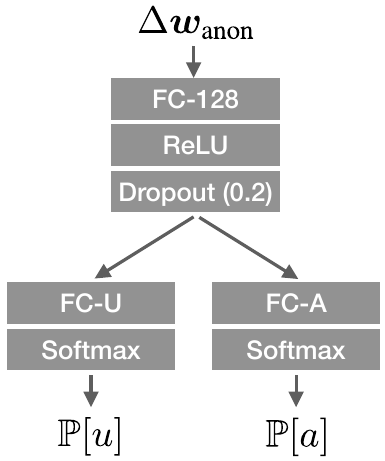}
  	\caption{MTL model to predict user $u$ and attribute $a$}
  	\label{fig:model_mtl}
  \end{subfigure}
  \caption{\textbf{Architectures of attack models}. Dotted lines indicate shared layers.}
\end{figure*}

To validate the assumption, we now present an experiment 
to quantify user variations on two public image datasets (PIPA \cite{piper} and OpenImages \cite{openimages}).
In both cases, we 
(i) group the images based on the real-world user who captured them using the corresponding \texttt{author} fields; and
(ii) vectorize images by extracting the 1024-dim \texttt{avgpool} features from MobileNet CNN \cite{howard2017mobilenets} and $L_2$-normalize them.
We obtain statistics for each user by computing two $L_2$ distances:
(a) intra-user distance: median image feature distance between images within each user; and
(b) inter-user distance: median image distance between user images and a set of random images.
We plot these distances per user on a scatter plot in Figure \ref{fig:dataset_distances}, each point indicating a distinct user.
If images captured by the users were unbiased, we would have found their corresponding points at the intersection of blue dashed lines.
However, points predominantly being above the diagonal indicates that examples within each users' collection are similar (low intra-user distances), but are greater (high inter-user distances) when compared to other user collections.
In Section \ref{sec:eval_reason_biases}, we further analyze how similar user-specific variations also arise in the parameter delta space.

The resulting non-IID distribution of user data $\data_u$ among devices 
leads to each device fitting a \textit{biased} estimator during the \texttt{DeviceUpdate} step (Algo. \ref{algo:federated}) with a bias error:
$\text{Bias}[\vecw_u] = \expect[\vecw_u] - \vecw^*$,
where the expectation term is over the user's training data $\data_u$ and $\vecw^*$ is the optimal estimator.
We conjecture (validated in \S\ref{sec:eval_reason_biases}) that the bias error signal is consistently encoded in both:
(i) the parameter updates transmitted by user's device $\Delta\vecw_u^t$; and
(ii) when estimating on prior data of the user $\vecw_u^\text{prior} = \text{SGD}(\data_u^\text{prior})$.
Hence, we reformulate the threat model (Eq. \ref{eq:threat1}) in the parameter update space:
\begin{equation}
    \label{eq:threat2}
    f^{\text{adv}}: \Delta \vecw_u^\text{prior} \times \Delta \vecw^t_{\anon} \rightarrow u \stackrel{?}{=} \anon
\end{equation}
Next, we look at attack models to learn this mapping.

\subsection{Attacks}
\label{sec:threat_attacks}

In this section, we present attack models to deanonymize users based on their model updates (Eq. \ref{eq:threat2}).

\myparagraph{Re-identification Attack}
In the re-identification scenario, the adversary leverages prior data to learn before-hand (via attack model $f^{\text{re-id}}$) what updates from targeted users look like.
The adversary then uses the attack model to re-identify users based on their anonymous update.
Formally, the re-identification attack involves training an attack model $f^{\text{re-id}}: \Delta\vecw_{u}^{\text{prior}} \rightarrow u$ to capture user-specific bias signals in the high-dimensional parameter delta space.
At test-time, users are re-identified using their model updates:
\begin{equation}
    \label{eq:reid_test}
	f^{\text{re-id}}: \Delta\vecw_{\anon} \rightarrow u
\end{equation}
For the re-identification attack model $f^{\text{re-id}}$, we adopt a Multilayer Perceptron (MLP) classifier (architecture in Fig. \ref{fig:model_mlp}) with a single hidden layer of 128 units and ReLU activation, trained using SGD with learning rate (LR) 0.01, 0.9 momentum and $10^{-6}$ LR decay.

\myparagraph{Matching Attack}
Instead of learning an update-to-user mapping, the adversary in the matching scenario learns a metric space among model updates.
Learning a metric space helps embed model updates close together if they are generated by the same user, independent of whether the user is a part of the adversary's prior knowledge base.
Formally, the adversary's objective is to predict the match probability of a pair of distinct parameter updates:
\begin{equation}
    \label{eq:matching}
    f^{\text{mat}}: (\Delta\vecw_i,\ \Delta\vecw_j) \rightarrow i \stackrel{?}{=} j
\end{equation}
where one or both parameter updates are anonymous.
The matching attack is particular helpful in scenarios where the adversary encounters novel users at test-time (\S \ref{sec:eval_attack_scenario}), or extending to cross-modal situations (discussed in next paragraph).
We adopt a Siamese network \cite{bromley1994signature} %
with metric learning \cite{weinberger2006distance} to perform the matching attack.
A Siamese model is characterized by twin networks which accepts distinct inputs ($\Delta\vecw_i$ and $\Delta\vecw_j$ in our case) and is connected by another network to estimate similarity between the individual embeddings produced by the twin networks.
In addition, the weights of the twin networks are shared to ensure extremely similar inputs are not mapped to distant embeddings. 
Our Siamese network (architecture in Fig. \ref{fig:model_siamese}) is constructed as :
(a) two FC-128 layers with ReLU activations which individually encodes $\Delta\vecw_i, \Delta\vecw_j$ into a 128-dim embedding;
(b) $L_1$ distance layer to represent distance between these embeddings; and
(c) FC-1 layer with sigmoid activation to predict the match probability. 
We minimize the binary-cross entropy loss and perform optimization using RMSProp with learning rate $10^{-3}$.

\myparagraph{Cross-modal Matching Attack}
We extend the matching attack to accommodate the situation where the modality of attacker's prior knowledge (e.g., text) differs from the private data (e.g., visual data) used by the users during training.
In such a scenario, parameter updates can no longer be represented in the same space (as in Eq. \ref{eq:threat2},\ref{eq:matching}).
As a result, the cross-modal matching attack performs:
\begin{equation}
    \label{eq:matching_cm}
    f^{\text{cm-mat}}: (\Delta\vecw_{\anon},\ \vec{\phi}_u) \rightarrow \text{anon} \stackrel{?}{=} u
\end{equation}
where $\vec{\phi}_u \in \real^D$ denotes an embedding of the user's prior data $\data_u^\text{prior}$.
In \S \ref{sec:eval_attack_priordist}, we discuss exactly how we obtain such an embedding.
The attack model (architecture in Fig. \ref{fig:model_crossmodal}) to estimate the match probability closely resembles the Siamese network for the matching attack.
The only modification is replacing the twin networks with two different networks (each with a single FC-128 layer) to map the inputs into a common 128-dim feature space.

\section{Experimental Setup: Datasets, Tasks, and Models}
\label{sec:dataset}
In this section, we discuss the experimental setup and datasets (summarized in Table~\ref{tab:datasets}) used to train and evaluate the collaboratively learnt ML model in an FL setup.

\begin{table*}[]
  \small
  \centering
  \begin{tabular}{@{}llllllll@{}}
  \toprule
  Dataset ($\data$)                                        & Type     & Task                       & \# Users & \# Examples & Input $(\dataspace)$ & Output $(\labelspace)$ & Model ($f_{\vecw}$)      \\ \midrule
  PIPA \cite{piper}             & Visual   & Multi-label classification & 53       & 33,051      & Image                & Labels                 & \modelname{CNN-PIPA-FL} \\
  OpenImages \cite{openimages}  & Visual   & Multi-label classification & 327      & 317,008     & Image                & Labels                 & \modelname{CNN-OI-FL}   \\
  Blog \cite{schler2006effects} & Language & Language Modeling          & 55       & 454,090     & Text                 & Text                   & \modelname{NNLM-FL}     \\
  Yelp \cite{challenge2013yelp} & Language & Sentiment Analysis         & 118      & 85,615      & Text                 & Score                  & \modelname{NNSA-FL}     \\ \bottomrule
  \end{tabular}
  \caption{\textbf{Datasets $\data$ and Models $f_{\vecw}$}. List of datasets used along with corresponding statistics, tasks, and models}
  \label{tab:datasets}
\end{table*}

\subsection{Datasets}
\label{sec:setup_datasets}

\begin{figure}[t]
  \centering
  \includegraphics[width=0.9\linewidth]{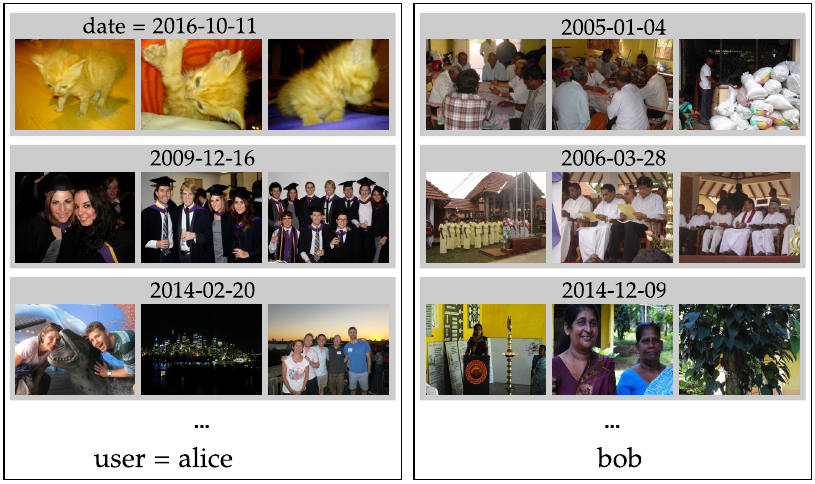}
  \vspace{0.5em}
  \includegraphics[width=0.9\linewidth]{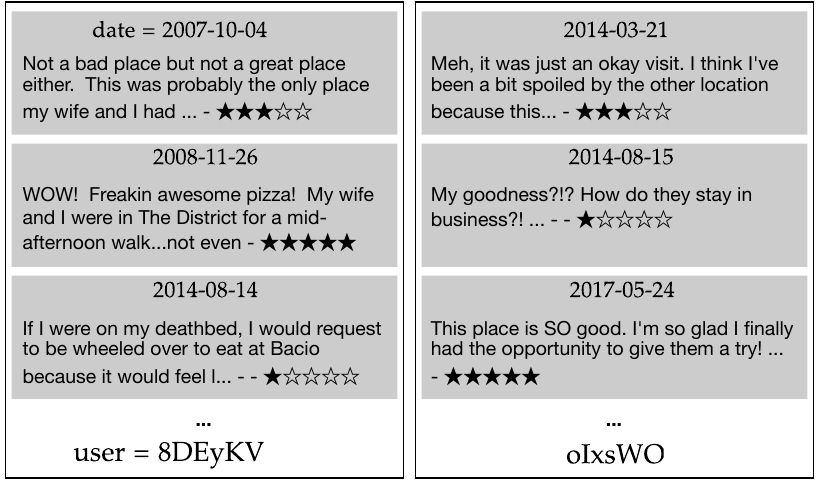}
  \caption{\textbf{Examples of users and corresponding data.} OpenImages (top) and Yelp (bottom). Images here are grouped by the anonymized userid and captured/review date.
  Qualitatively, we observe that the difference between users' data is typically subtle.}
  \label{fig:openimages_example}
\end{figure}

We now present the datasets (Table~\ref{tab:datasets}, examples in Fig.~\ref{fig:openimages_example}) used to train and evaluate the collaboratively trained models $f_{\vecw}$.
We highlight that the datasets used are well-suited since:
(a) they are publicly available;
(b) samples are annotated with non-private labels (e.g., tv, flower);
(c) examples are complex and realistic; and
(d) each training example has a notion of ``owner'' or ``user''.
Property (d) is particularly important in FL scenarios, as it allows us to partition and distribute data on devices based on user identities.
Each of the following paragraphs discusses the 
(i) dataset $\data$; 
(ii) corresponding task $\dataspace \rightarrow \labelspace$; and
(iii) training model $f_{\vecw}: \dataspace \rightarrow \labelspace$ to perform the task.

\myparagraph{(i) PIPA}
PIPA \cite{piper} is a dataset consisting of $\sim$37k personal photos uploaded by actual Flickr users (indicated in the \texttt{author} field in Flickr photo metadata).
To assure certain minimal amount of per-user data, we only use users with at least 100 images, resulting in 33K images over 53 users.
We obtain labels for each image by running a state-of-the-art object detector \cite{huang2017speed} that detects 80 COCO \cite{lin2014microsoft} classes, such as umbrella, backpack, and bicycle.
To perform reasonable training and evaluation of the multilabel classification task, we use 19 classes (e.g., chair, cup, tv) that occur in approximately $>$1\% of images with high precision.
We train a multi-label image classifier \texttt{CNN-PIPA-FL} $f_{\vecw}: \mathbb{R}^{224 \times 224 \times 3} \rightarrow \real^{19}$, for this dataset in an FL setup.
We use the MobileNet \cite{howard2017mobilenets} architecture designed specifically to be run on mobile devices, as it is a lightweight architecture that strikes a good balance between latency, accuracy and size.

\myparagraph{(ii) OpenImages}
OpenImages \cite{openimages} is a large-scale public dataset from Google, consisting of 9M Flickr image URLs and weakly labeled image-level annotations across 19.8k classes. 
To make training feasible, we prune out users with less than 500 images,
resulting in 317k images from 327 users annotated with 18 classes (e.g., food, building).
Furthermore, images of the same user can cover a wide time span (typically >5 years).
Similar to PIPA, we formulate the training of a multi-label image classifier \modelname{CNN-OI-FL} based on the MobileNet architecture.

\myparagraph{(iii) Blog Authorship}
The Blog Authorship Corpus \cite{schler2006effects} contains $\sim$681K posts collected from 19K bloggers from \url{blogger.com}.
We work with a subset of 55 users with at least 1000 corresponding posts.
Since these blog posts are lengthy (13.5 sentences, 209 words per post), we further split each post into corresponding sentences.
As a result, we obtain 454K text sequences over 55 users.
We train a language model (\modelname{NNLM-FL}): $P(\vecx_t | \vecx_{t-i}, \cdots, \vecx_{t-1};\ \vecw)$ i.e., predicting probability distribution of the next word $\vecx_t$ in a sequence given contextual information.
Language models trained in an FL architecture are currently deployed to enable smart compose keyboards \cite{yang2018applied}.
We train a Neural Network Language Model \cite{bengio2003neural} using an embedding layer (with $E$=100 dims), LSTM layer \cite{hochreiter1997long} (with $L$=64 hidden units), and a fully-connected layer (with vocabulary size $V$=5000).

\myparagraph{(iv) Yelp}
The Yelp Dataset \cite{challenge2013yelp} contains $\sim$6M user-reviews of 188K businesses.
To allow for each user contributing meaningful parameter deltas, we filter users with at least 500 total reviews.
This results in 85K user reviews over 118 users.
Each user review contains text (mean length = 180 words) and a 1-5 star rating.
We train a sentiment analyzer, modeled as a neural network regressor: $y = f_{\vecw}([\vecx_1, \vecx_2, \cdots])$, where $y \in [1, 5]$ is the rating and $\vecx_i$ is a representation of $i$-th word in the review.
We use a standard recurrent neural network architecture with an embedding size of $E$=50, $L$=128 hidden LSTM units, and a vocabulary size of $V$=1000.

\subsection{Data Setup for Adversarial Knowledge}
\label{sec:threat_adv_know}

\label{sec:setup_adv_knowledge}
The datasets collected (Table~\ref{tab:datasets}) contain sets of user-specific data $\data_u = \{ (\vecx_i, y_i) \}_{i=1}^{n_u}$ over users $u \in \userset$.
A limited subset of this data is strategically held-out to model the adversary's prior knowledge $\data^{\text{prior}}_u$, and the remaining used as the users' private training data $\data^{\text{private}}_u$.
We consider multiple prior-data limitation strategies to systematically study their influence on deanonymization attacks:
(i) limiting the subset of users the adversary has prior knowledge on (\S \ref{sec:threat_scenarios}); and
(ii) limiting the amount and quality of prior knowledge (\S \ref{sec:threat_priordist}).

\subsubsection{User Scenarios}
\label{sec:threat_scenarios}
To tackle the case where a subset of participating users in FL may or may not be a part of adversary's prior knowledge database, we set-up two scenarios:

\myparagraph{Closed-world}
The adversary has some prior information on all users participating anonymously in FL. 
Consequently, deanonymization of a particular device always maps to a closed-set of `seen' users.
This scenario captures instances of silo-based federated learning scenarios, which typically involve a small number of organizations (the users).

\myparagraph{Open-world}
We extend the above world to additionally include `unseen' users during FL, for which the adversary does not have prior information. Hence, a parameter delta $\Delta\vecw_{\anon}$ could map either to a seen or an unseen user.
This presents a challenging scenario, as it leads to `finding a needle in a haystack' i.e, the adversary wants to re-identify a particular target user in spite of background noise generated by many unseen users.

\subsubsection{Type of Prior Knowledge}
\label{sec:threat_priordist}
To understand the role of prior information in a systematic manner, we consider both the \textit{amount} and \textit{distribution} of adversary's prior information w.r.t private data on the FL device.
Specifically for the distribution, we model both $\data^{\text{prior}}_u$ and $\data^{\text{private}}_u$ to be sampled (without replacement) from user $u$'s universal data distribution $\data_u$ in one of the four following manners.

\noindent \textbf{(i) \texttt{random} prior:}
Both the prior and private data are IID samples from $\data_u$  i.e., $\data^{\text{prior}}_u, \data^{\text{private}}_u \stackrel{\text{iid}}{\sim} \data_u$
This scenario captures the adversary scraping information on target user $u$ randomly from various social media sources.

\noindent \textbf{(ii) \texttt{chrono} prior:}
We also consider both prior and private data to be sampled non-IID from $\data_u$ by factoring in timestamps of data (e.g., from image EXIF metadata).
Here, data in $\data^{\text{prior}}_u$ chronologically precedes data in $\data^{\text{private}}_u$
For instance, this could occur when an adversary has historical data on the targeted user, such as from a previously de-identified account.
In the specific case of the PIPA dataset, where the exact timestamp per example is unavailable, we sample prior and private data non-IID using album information (\texttt{photoset} field).

\noindent \textbf{(iii) \texttt{profile} prior:}
We briefly address a scenario where the adversary uses a set of curated `profile' data as a proxy to users' data.
For instance, by curating targeted prior data $\data^{\text{prior}}_u$ to specifically contain weapons to identify participating users who fit that profile.

\noindent \textbf{(iv) \texttt{cross-model} prior:}
We consider the case where adversary's prior data  of the user $\data^{\text{prior}}_u$ is gathered from a different modality compared to the private
data. 
For instance, where the prior data is text-based, but the users train on visual data.

\begin{table}[]
  \scriptsize
  \centering
  \begin{tabular}{@{}ccc@{}}
      \toprule
                  & \multicolumn{2}{c}{PIPA} \\  \cmidrule{2-3}
      split   & \texttt{random}      & \texttt{chrono}     \\ \midrule
      \modelname{CNN-PIPA-FL} & 45.1       & 37.7        \\
      \modelname{CNN-PIPA-SGD}   & 49.7       & 40.7        \\
      K-NN        & 14.9       & 15.8        \\
      Chance      & 9.5        & 9.7         \\ \bottomrule \\
  \end{tabular}
  \begin{tabular}{@{}ccc@{}}
    \toprule
                & \multicolumn{2}{c}{OpenImages} \\  \cmidrule{2-3}
      split   & \texttt{random}      & \texttt{chrono}     \\ \midrule
    \modelname{CNN-OI-FL} & 62.9          & 62.2           \\
    \modelname{CNN-OI-SGD}   & 68.0          & 67.8           \\
    K-NN        & 9.7           & 13.6           \\
    Chance      & 6.3           & 6.3            \\ \bottomrule \\
    \end{tabular}
    \begin{tabular}{@{}ccc@{}}
      \toprule
              & \multicolumn{2}{c}{Blog} \\ \cmidrule{2-3}
      split   & \texttt{random}      & \texttt{chrono}     \\ \midrule
      \modelname{NNLM-FL}  & 28.02       & 27.83      \\
      \modelname{NNLM-SGD}    & 28.62       & 28.22      \\
      Chance  & 0.09        & 0.09       \\ \bottomrule
  \end{tabular}
  \qquad \quad
  \begin{tabular}{@{}ccc@{}}
    \toprule
           & \multicolumn{2}{c}{Yelp} \\ \cmidrule{2-3}
    split   & \texttt{random}      & \texttt{chrono}     \\ \midrule
    \modelname{NNSA-FL} & 0.716       & 0.708      \\
    \modelname{NNSA-SGD}   & 0.576       & 0.602      \\
    Chance & 1.472       & 1.514      \\ \bottomrule
  \end{tabular}
    \caption{\textbf{Evaluation of $f_{\vecw}$.} Datasets from Table \ref{tab:datasets}. Metrics used are:
    (a) PIPA: Average Precision (AP)
    (b) OpenImages: Average Precision (AP)
    (c) Blog: Top5 accuracy
    (d) Yelp: Mean Absolute Error (MAE).
    For (a-c), higher is better and for (d), lower is better.
    }
  \label{tab:collab_perf}
\end{table}

\subsection{Collaborative Models: Training and Performance}
\label{sec:setup_collab}

In Section \ref{sec:setup_datasets}, we discussed details on the datasets and corresponding model architectures $f_{\vecw}$.
Section \ref{sec:setup_adv_knowledge} presented how we strategically hold-out a subset of the data to serve as adversary's prior knowledge.
Now we discuss setup and performances of collaborative models in our FL setting.

\myparagraph{Training Models $f_{\vecw}$}
For each dataset, we train models $f_{\vecw}$ using \texttt{FederatedAveraging} (Algorithm \ref{algo:federated}) \cite{McMahan2017CommunicationEfficientLO}.
For all models, crucial hyper-parameters (e.g., size of vocabulary or embedding) were selected carefully after rigorous evaluation over a set of standard choices.
In \texttt{FederatedAveraging} algorithm, we use $C$=0.1 and $E$=1, which we empirically find results in a good trade-off between convergence and communications required.
We train the models for 200 epochs with learning rate $\eta$=0.01, resulting in 1-4 GPU days to train a single model for a particular architecture, dataset and scenario.
All models are written in Python using the Keras \cite{chollet2015keras} library with a TensorFlow \cite{abadi2016tensorflow} back-end.

\begin{table*}[t]
  \footnotesize
  \centering
  \begin{tabular}{@{}ccccccccccccccccc@{}}
    \toprule
           &  & \multicolumn{7}{c}{PIPA (\#Users $U$ = 53)}                                                 &  & \multicolumn{7}{c}{OpenImages ($U$ = 327)}                                           \\ \cmidrule{3-9} \cmidrule{11-17}
           &  & \multicolumn{3}{c}{\texttt{random}}        &  & \multicolumn{3}{c}{\texttt{chrono}}      &  & \multicolumn{3}{c}{\texttt{random}}        &  & \multicolumn{3}{c}{\texttt{chrono}}        \\ \cmidrule{3-5} \cmidrule{7-9} \cmidrule{11-13} \cmidrule{15-17}
           &  & AP                & Top-1 & Top-5 &  & AP                & Top-1 & Top-5 &  & AP                & Top-1 & Top-5 &  & AP                & Top-1 & Top-5 \\ \midrule
    MLP    &  & 91.0 (48$\times$)   & 84.7  & 96.3  &  & 42.2 (22$\times$) & 40.0    & 68.8  &  & 53.7 (175$\times$) & 51.9  & 77.9  &  & 32.5 (106$\times$) & 31.9  & 57.1  \\
    SVM    &  & 81.3 (43$\times$) & 89.3  & 91.9  &  & 27.7 (15$\times$) & 43.7  & 49.6  &  & 49.0 (159$\times$)   & 66.5  & 67.0    &  & 24.6 (80$\times$) & 41.7  & 42.5  \\
    kNN    &  & 85.4 (45$\times$) & 82.6  & 92.6  &  & 31.5 (17$\times$) & 38.4  & 54.8  &  & 46.0 (150$\times$)   & 49.2  & 63.9  &  & 25.1 (82$\times$) & 30.3  & 43.1  \\
    Chance &  & 1.9 (1$\times$)  & 2.0     & 9.9   &  & 1.9 (1$\times$)  & 2.0     & 9.9   &  & 0.3 (1$\times$)  & 0.3   & 1.5   &  & 0.3 (1$\times$)  & 0.3   & 1.5   \\ \bottomrule
  \end{tabular} 
  
  	\vspace{1.0em}
    \begin{tabular}{@{}ccccccccccccccccc@{}}
    \toprule
           &  & \multicolumn{7}{c}{Blog ($U$ = 55)}                                                 &  & \multicolumn{7}{c}{Yelp ($U$ = 118)}                                           \\ \cmidrule{3-9} \cmidrule{11-17}
           &  & \multicolumn{3}{c}{\texttt{random}}        &  & \multicolumn{3}{c}{\texttt{chrono}}      &  & \multicolumn{3}{c}{\texttt{random}}        &  & \multicolumn{3}{c}{\texttt{chrono}}        \\ \cmidrule{3-5} \cmidrule{7-9} \cmidrule{11-13} \cmidrule{15-17}
           &  & AP                & Top-1 & Top-5 &  & AP                & Top-1 & Top-5 &  & AP                & Top-1 & Top-5 &  & AP                & Top-1 & Top-5 \\ \midrule
      MLP                                 &  & 52.9 (29$\times$)  & 50.1 & 89.9 &  & 44.8 (25$\times$)  & 47.6 & 81.3 &  & 23.5 (28$\times$)  & 25.2 & 50.1 &  & 16.0 (19$\times$) &   18.9    &  38.9     \\
  SVM                                 &  & 35.7 (20$\times$)  & 46.3 & 49.2 &  & 27.0 (15$\times$)  & 42.1 & 46.0 &  & 25.9 (31$\times$)  & 43.2 & 44.9 &  & 17.1 (20$\times$) &  33.3     &  36.7     \\
  kNN                                 &  & 35.6 (20$\times$)  & 39.8 & 64.9 &  & 29.5 (16$\times$)  & 35.6 & 58.3 &  & 21.6 (25$\times$)  & 25.3 & 41.1 &  & 15.4 (18$\times$) &  21.0     &  32.9     \\
  Chance                              &  & 1.8 (1$\times$)  & 1.7 & 8.8 &  & 1.8 (1$\times$)  & 1.6 & 8.8 &  & 0.9 (1$\times$)  & 0.8 & 4.1 &  & 0.9 (1$\times$) &  0.9     &  4.3     \\ \bottomrule
  \end{tabular}
  \caption{\textbf{Re-identification Attack Evaluation ($\Delta\vecw_\anon \rightarrow u$).} Performed in a closed-world. Chance-level AP $\approx$ 1/$U$.
  }
  \label{tab:eval_identification_closed}
\end{table*}

Each user $u$ in our datasets is associated with a variable number of examples $\data_u$ sampled according to some distribution (e.g., \texttt{chrono}; see \S \ref{sec:threat_scenarios}).
By default, we place half of the users' data $\data_u$ on their anonymous device and reserve the remaining to be used as adversary's prior knowledge.
In Section \ref{sec:eval_attack_priornum}, we vary the size of the adversary's prior knowledge and find attacks possible even in severely data-limited settings (e.g., 1-50 prior samples).

\myparagraph{Evaluation of $f_{\vecw}$}
We evaluate performance of the collaboratively-trained models on a 20\% held-out test set.
For reference, we similarly evaluate models  trained in a centralized manner i.e., standard training from a single pool of training data.
The performances of FL-trained models (represented as `\modelname{X-FL}') and SGD-trained models (`\modelname{X-SGD}') are presented in Table \ref{tab:collab_perf}.
When possible, we also present the $K$-Nearest Neighbours (KNN, with $K$=10) baseline.
We observe strong performances of the FL-trained models $f_{\vecw}$ across all datasets, where they consistency recover $80-98$\% performance of models trained using centralized SGD.

\section{Evaluation}
\label{sec:eval}
In the previous section, we discussed training ML models in an FL setup for four different datasets covering various tasks such as image classification and language modeling.
Within this FL scenario, we now detail the training of \textit{deanonymization attack} models (\S \ref{sec:threat_attacks}), evaluate their effectiveness, and work towards understanding how the parameter updates leak user-identifiable information.

\myparagraph{Evaluation Metrics}
We use the following metrics (computed using scikit-learn \cite{scikit-learn}) to evaluate the adversary's attack performance:
\begin{enumerate*}[label=(\roman*)]
	\item \textbf{Mean Average Precision (AP)}: Adversary's precision-recall curves for held-out user data is computed. We then compute the per-user Average Precision (area under the precision-recall curves). We report the mean of Average Precisions across users in percentages (i.e., AP$\times$100);
	\item \textbf{Increase over Chance}: In order to analyze adversary's information gain, we compute this as (predicted AP)/(chance AP). We display this alongside AP scores in the form: $\Box\times$; and
	\item \textbf{Top-1 accuracy}: We compute the classification success rates over all parameter updates in the test set.
\end{enumerate*}
These metrics are common among classification tasks e.g., \cite{everingham2010pascal,lin2014microsoft,wang2016cnn} for AP and \cite{krizhevsky2012imagenet,he2016deep,deng2009imagenet} for Top-1 accuracy.
We use the AP as the primary metric, since it also takes into account ranking among predicted classes.

\myparagraph{Training and Evaluation Data for Attacker $f^{\text{adv}}$}
We train the ML models ($f_{\vecw}$ in Table \ref{tab:datasets}) in an FL system simultaneously using two disjoint sets of devices per user: 
(a) $\clientset_{\text{anon}}$: anonymous user devices (that adversary wants to deanonymize); and
(b) $\clientset_{\text{prior}}$: adversary's shadow devices containing target users' prior information (that we use to generate training data for attack models in \S \ref{sec:threat_attacks}).
For simplicity, we restrict each of these sets to contain a single user.
During training of $f_{\vecw}$ over multiple rounds, we accumulate the parameter updates $\Delta \vecw_k^t$ communicated by all devices in FL.
To train the attack models $f^{\text{adv}}$, we use the set of parameter updates $\{ (\Delta\vecw_k^t, u) : k \in \clientset_{\text{prior}}\}$, where we know a priori the device $k$ to user $u$ mapping.
We discuss in detail training data-limited adversaries in Section \ref{sec:eval_attack_priornum}.
We evaluate attacks on the disjoint set of parameter updates $\{\Delta (\vecw_k^t, u) : k \in \clientset_{\text{anon}}\}$.

\myparagraph{Representing $\Delta\vecw_k^t$ for Attacks} 
The parameter updates contain hundred thousands to millions of parameters.
To enable faster training and evaluation of attack models, we choose a subset of parameters by representing $\Delta\vecw_k^t$ using weights of layers which achieves best attack performance:
\begin{enumerate*}[label=(\roman*)]
	\item \modelname{CNN-PIPA-FL, CNN-OI-FL}: Fully Connected Layer (19K parameters);
	\item \modelname{NNLM-FL}: LSTM layer (10K parameters); and
	\item \modelname{NNSA-FL}: Embedding layer (50K parameters).
\end{enumerate*}
This has little impact to our attack; influence of each layer is discussed in Section \ref{sec:eval_ext_layer}.
Furthermore, we flatten $\Delta\vecw_k^t$ into a vector and $L_2$ normalize it.

\subsection{Effectiveness of Deanonymization Attacks}
\label{sec:eval_attack}

In this section, we validate effectiveness of the deanonymization attacks.
We begin by understanding the effectiveness in relation to adversary's prior knowledge (\S \ref{sec:eval_attack_priordist} and \S \ref{sec:eval_attack_scenario}) and discuss how it can be coupled with attribute inference attacks (\S \ref{sec:eval_att_inf}).

\subsubsection{Impact of Adversary's Prior Distributions}
\label{sec:eval_attack_priordist}

In this section, we focus on how \textit{types} of adversary's prior knowledge (\S \ref{sec:threat_priordist}) influences effectiveness of deanonymization.
Consequently, we address a range of scenarios, such as when the adversary has similar (\texttt{random}) or historical prior data (\texttt{chrono}) of the targeted users to perform deanonymization.
We also evaluate the novel challenge where the prior data is from a different modality (\texttt{cross-modal}).

\myparagraph{Leveraging \texttt{random} and \texttt{chrono} prior to deanonymize}
We present key results of the re-identification attack model `MLP' (\S \ref{sec:threat_attacks}): $f^{\text{re-id}}: \Delta \vecw_{\text{anon}}^t \rightarrow u$ (in a closed-world setting) in Table \ref{tab:eval_identification_closed}.
In addition, as baseline attack methods, we also demonstrate performances of `SVM' (a linear support vector machine) and `kNN' (a $k$-nearest neighour classifier using $k$=10).

From the results presented in Table \ref{tab:eval_identification_closed}, we observe:
\begin{enumerate*}[label=(\roman*)]
  \item All deanonymization attacks greatly outperform chance-level performances, with as much as 175$\times$ boost for MLP on the OpenImages dataset under the \texttt{random} prior, highlighting the effectiveness of the proposed deanonymization attack;
  \item Even the most simple K-NN attack is reasonably effective and already presents a significant threat (150$\times$ over random chance on OpenImages, \texttt{random} prior);
  \item MLP is highly effective across all datasets and splits (175$\times$ over random chance on OpenImages, \texttt{random} prior);
  \item Although the absolute AP scores are lower for the more challenging and larger OpenImages dataset (53.7\% AP on \texttt{random} prior), the increase over chance level performance is significantly higher (48$\times$ on PIPA vs. 175$\times$ on OpenImages under the same \texttt{random} prior);
  \item The attack is effective (19-106$\times$) even on \texttt{chrono} priors, where the adversary uses historical prior information to deanonymize users.
\end{enumerate*}

The above experiments were performed in a non-IID data-distribution among devices, which is natural in FL since users participate with personal data exhibiting unique biases (\S \ref{sec:threat_bias}).
We also perform attack evaluation in a contrasting IID setup, where we manually unbias data on devices by replacing each user example with an example drawn IID from $\data = \bigcup_k \data_k$.
We observed near-chance-level adversary performance (e.g., 1.5$\times$ chance-level for PIPA) since user data is no longer characteristic.
There is strong evidence that anonymous model parameter updates contain ample user information in an FL setup that allows for effective deanonymization.

\myparagraph{Cross-modal attacks}
We now evaluate the effectiveness of deanonymization attacks with a \texttt{cross-modal} prior (Section \ref{sec:threat_priordist}).
Here, the adversary is limited to prior knowledge from a \textit{different} modality from the data used during training by the users.
In particular, we consider the case where the prior data consists of text samples and the private data consists of images.
As we are not aware of any dataset which provides cross-modal user-generated data to evaluate the attack, we substitute PIPA prior image samples with corresponding text-representations obtained using a Neural Image Caption generator \cite{vinyals2015show}.
Using this setup, we train the cross-modal matching network $f^{\text{cm-mat}}: (\Delta\vecw_{\anon},\ \vec{\phi}_u) \rightarrow \text{anon} \stackrel{?}{=} u$ (Eq. \ref{eq:matching_cm}).
To obtain a compact text representation $\vec{\phi}_u$ over the prior knowledge (set of text sentences for a particular user), we:
(i) obtain the 4096-dim sentence-level embedding using InferSent \cite{conneau2017supervised}; and
(ii) compute the mean over the sentence embeddings for the user.
We evaluate $f^{\text{cm-mat}}$ on a balanced set of 10K pairs $\{((\Delta\vecw_{\anon},\ \vec{\phi}_u), \mathbbm{1}_{{\anon}=u})\}$.
We observe an attack performance of 76.3 AP (chance = 50.0 AP), indicating that model updates can be interestingly deanonymized even using data from another modality.

\begin{figure}
    \centering
  \includegraphics[width=0.9\linewidth]{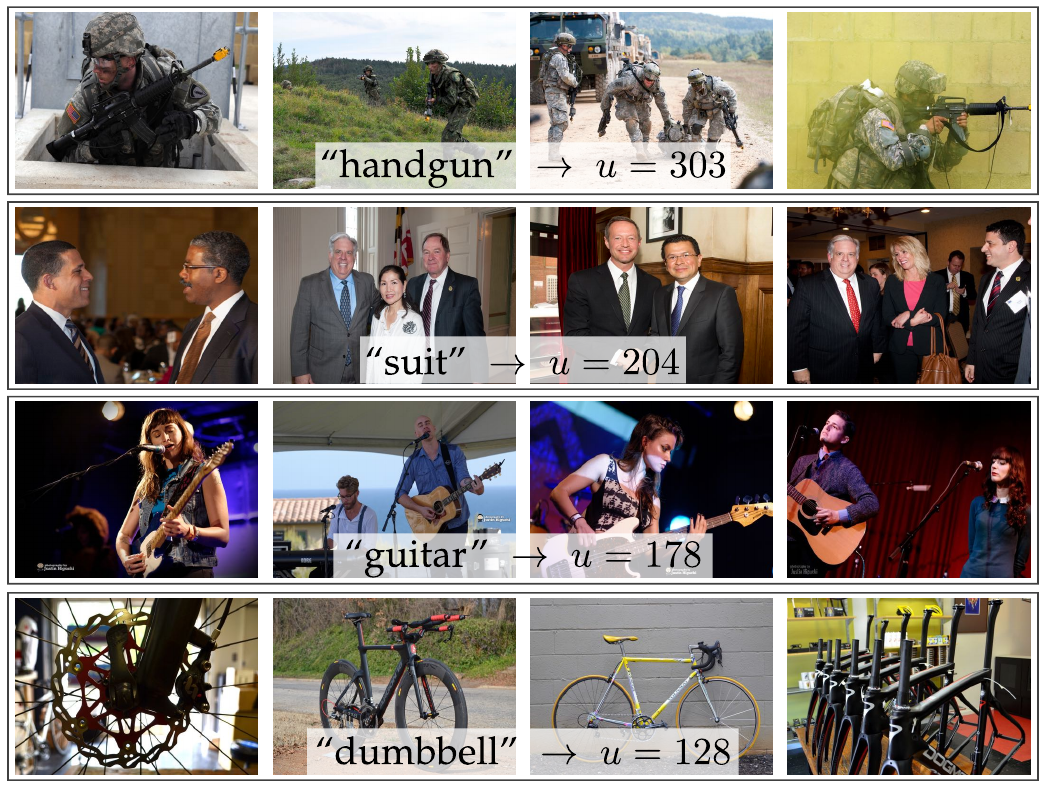}
  \caption{\textbf{\texttt{profile} prior.} Devices can be isolated using proxy distributions of certain profiles e.g., guitars. Rows denote private data $\data^{\text{private}}_u$ of users on devices.} %
  \label{fig:qual_profile}
\end{figure}

\myparagraph{Attacking using \texttt{profile} prior}
In the previous attacks we looked at the task of deanonymizing devices by associating the parameter updates to prior data of users.
We now look at a slightly different task of linking devices that fit a certain \texttt{profile} prior.
We achieve this by manually constructing $\data^\text{profile}$ to comprise of examples of interest e.g., weapons.
In Figure \ref{fig:qual_profile} we display the top users (in the OpenImages dataset) found using the re-identification attack who fit the corresponding profiles.
We observe:
(i) devices can be remarkably singled out using various proxy distributions (of e.g., handgun, guitar) circumventing the need for real user data; 
(ii) however, valid correlations in data can sometimes lead to false positives. For instance, `dumbbells' which often co-occur in images along with other physical equipment devices leads to bicycle images of user 128 (which also displays similar correlations) being falsely identified.

\subsubsection{Impact of Number of Seen and Unseen Users}
\label{sec:eval_attack_scenario}

In the previous section, we evaluated attacks in a closed-world scenario (\S \ref{sec:threat_scenarios}), where the adversary was aware of every users' existence (i.e., included in prior knowledge).
We now consider the open-world scenario, where at test-time the adversary additionally encounters model updates generated by \textit{unseen} users (i.e., not in the prior knowledge).
This introduces the challenge of differentiating between seen and unseen identities when deanonymizing.

\myparagraph{User Split}
In our experimental setup, we split the users $\userset$ into three variably-sized disjoint sets:
(a) $\userset_{\text{unseen}}$: prior data is unavailable and should be classified as unseen at test-time;
(b) $\userset_{\text{seen}}$: prior data is available and should be deanonymized at test-time; and
(c) $\userset_\text{holdout}$: these users are reserved purely for training purposes.

\myparagraph{Re-identification Setup}
Previously in the closed-world scenario, we trained the MLP (\S\ref{sec:threat_attacks}) classifier $f^{\text{re-id}}: \Delta\vecw_k \rightarrow u$ with $|\userset|$ classes representing all users at test time.
Now we train a similar classifier over $|\userset_{\text{seen}}| + 1$ output classes with the additional class \texttt{unseen} collectively denoting unseen users.
During training, we use users $\userset_\text{holdout}$ and their parameter updates to train the \texttt{unseen} class.

\begin{figure}[t]
    \centering
  \includegraphics[width=\linewidth]{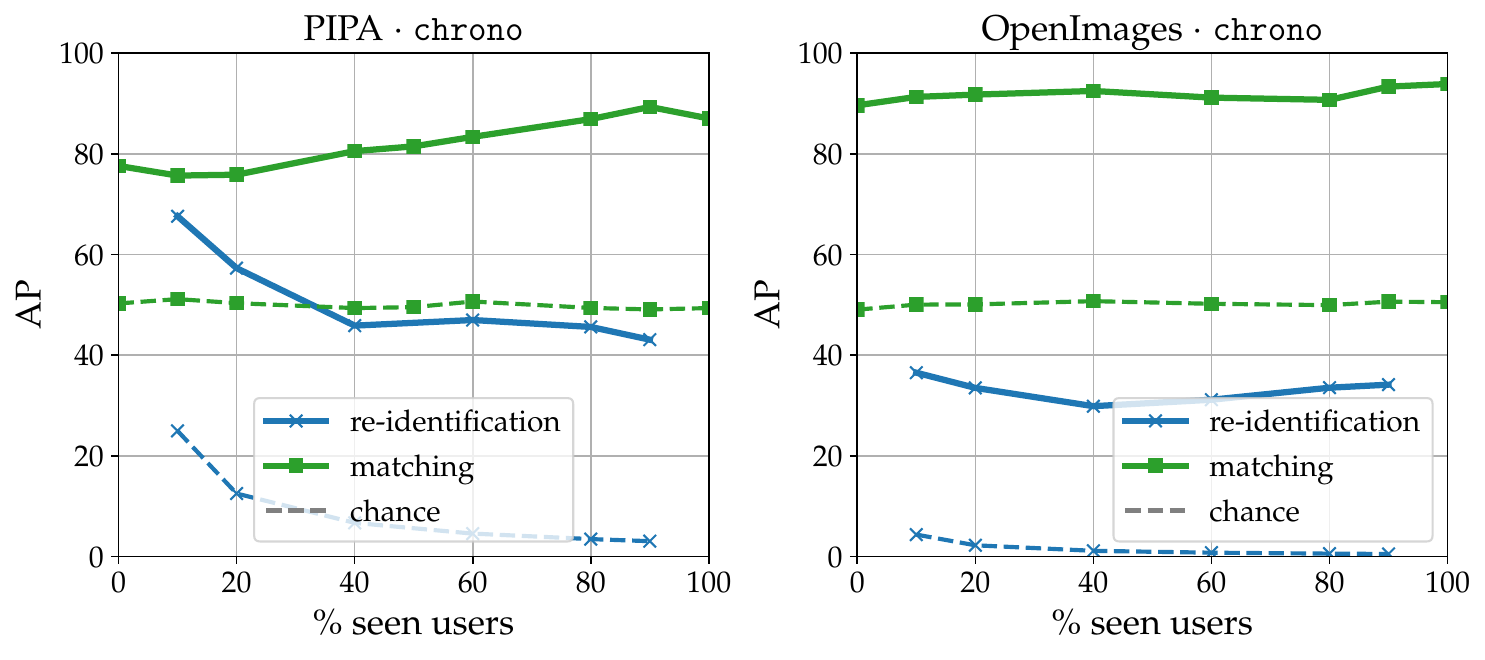}
  \includegraphics[width=\linewidth]{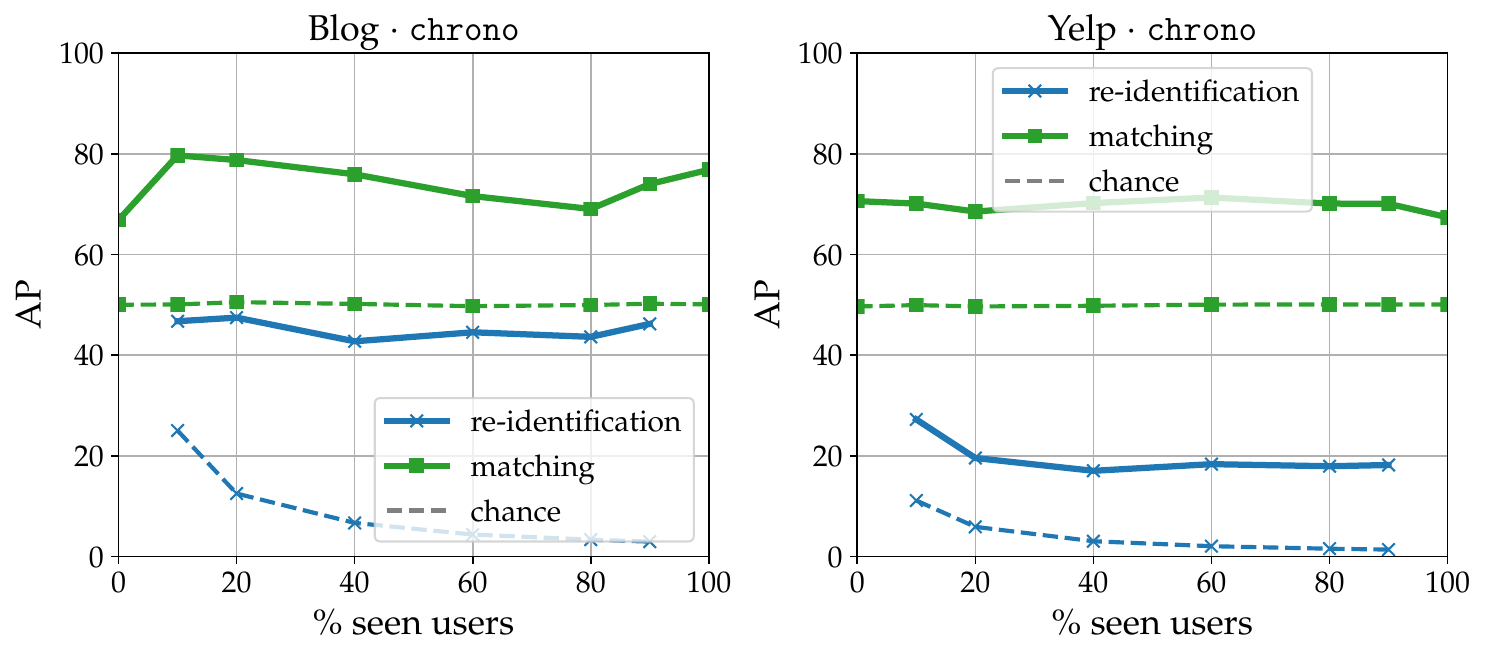}
  \caption{\textbf{Open-world evaluation.} Across re-identification (MLP) and matching (Siamese) attack models.} %
  \label{fig:openworld_clf_linking}
\end{figure}

\myparagraph{Matching Setup}
We train a Siamese network (\S\ref{sec:threat_attacks}) using parameter updates from  held-out and seen set of users.
Given a pair $(\Delta\vecw_i, \Delta\vecw_j)$, the network predicts the probability $\mathbb{P}[i = j]$ of being generated by the same user.

\myparagraph{Evaluation}
The performances are evaluated at different ratios of seen and unseen users at test time.
We keep the size of the hold-out set constant to one-third of the total number of users.
Evaluation for both re-identification and matching tasks on the challenging \texttt{chrono} prior distributions per dataset are presented in Figure \ref{fig:openworld_clf_linking}.
We observe:
\begin{enumerate*}[label=(\roman*)]
  \item even in the open-world scenario, we perform much higher than chance-level for both the tasks 
  consistently across a wide range of seen vs. unseen scenarios;
  \item for the re-identification attack, as \% seen users increase, the complexity of the task increases as well (due to larger output-space). Hence,
  we notice a drop in AP performance (67\%$\rightarrow$43\% in PIPA). However, performance compared to chance-level significantly increases (3$\times\rightarrow$14$\times$);
  \item in the matching task, the Siamese model performs much higher than chance-level even in a purely open-world setting, with no seen users (1.5$\times$ for PIPA and 1.8$\times$ for OpenImages).
\end{enumerate*}
We find both the re-identification and matching attacks generalize well in the presence of unseen users at test time.

\subsubsection{Amplification with Attribute Inference Attacks}
\label{sec:eval_att_inf}

We now discuss how deanonymization attacks can be coupled with related inference attacks on model updates.
Specifically, we consider the recent attribute inference attack \cite{melis2018inference}, which recovers sensitive properties (e.g., race) that holds for subsets of training data. 
In this particular case, our attack objective involves jointly inferring both identity (via our deanonymization attacks) and sensitive attributes (via attribute inference attacks) via transmitted model updates.

\begin{table}[t]
\centering
\footnotesize
\begin{tabular}{@{}lllllll@{}}
\toprule
 &             & \multicolumn{2}{c}{STL} &  & \multicolumn{2}{c}{MTL} \\ \cmidrule{3-4} \cmidrule{6-7}
Attributes    & \# Attrs & AttrInf                 & Deanon                &  & AttrInf                & Deanon                \\ \midrule
Age            & 5           & 89.1                   & -                  &  & 90.8                  & 90.9                  \\
Gender         & 2           & 93.1                   & -                  &  & 94.4                  & 91.6                  \\
Glasses        & 3           & 98.5                   & -                  &  & 98.9                 & 91.3                  \\
Hair Color     & 3           & 85.2                   & -                  &  & 88.7                  & 90.1                  \\
Hair Length    & 5           & 91.3                   & -                  &  & 91.3                  & 90.1                  \\ 
-            & -           & -                   & 87.6                  &  & -                  & -                  \\ \bottomrule
\end{tabular}
\caption{\textbf{Attribute Inference and Deanonymization Attack Performances.} Results are reported in top-1 accuracies. Columns indicate when the inference tasks are trained individually (STL) and jointly (MTL). %
}
\label{tab:att_inf}
\end{table}

To evaluate the attacks, we closely follow the data setup on Melis et al. \cite{melis2018inference} on the PIPA dataset.
Attribute inference in this setting involves inferring sensitive attributes (e.g., age) from the model updates.
To this end, we first train individual attribute classification models for each of the five attributes, and an additional re-identification model.
All the classification models are MLPs following the architecture of the re-identification model.
Table \ref{tab:att_inf} (column STL) presents results over the five attribute inference (column AttrInf) tasks and deanonymization (column Deanon).
Here, we observe that an attacker can consistently achieve 85.2-98.5\% accuracy in inferring various attributes from model updates and 87.6\% accuracy in inferring identities of participants.
These results suggest that model updates indeed leak details unrelated to the trained task (recognizing chair, couch, etc.) and allows an attacker to recover sensitive attributes of the users' training data (via attribute inference) and further link them to an identity (via deanonymization).

We now recast the problem of inference on attributes and identities as a multi-task learning (MTL) \cite{caruana1997multitask} problem.
The core idea is to exploit commonalities between the two related tasks to learn a better representation jointly benefiting the tasks.
To achieve this, we extend our re-identification model (\S \ref{sec:threat_attacks}, which performs user classification) with a secondary classification head (which performs attribute inference; see Fig. \ref{fig:model_mtl}).
Consequently, the model is simultaneously trained for both attribute inference and deanonymization using their corresponding losses.
The results for the model is presented under the MTL column in Table \ref{tab:att_inf}.
We observe by learning the two tasks jointly can improve attribute inference performances consistently by 0-3.5\% and deanonymization by 2.5-4\%.
Our results suggest that apart from jointly inferring sensitive attributes and recovering identities, the two related attacks surprisingly amplify each other's performances.

\subsection{Analysis}
\label{sec:eval_analysis}
In this section, we take a closer look at various factors that influence (e.g., amount of training data) the effectiveness of attacks.
For simplicity, we study the factors using the re-identification attack in a closed-world setup.
We conclude the section by reasoning why model updates lend themselves to deanonymization risks.

\subsubsection{Amount of Training Data}
\label{sec:eval_attack_priornum}
\begin{figure}[t]
  \centering
  \includegraphics[width=\linewidth]{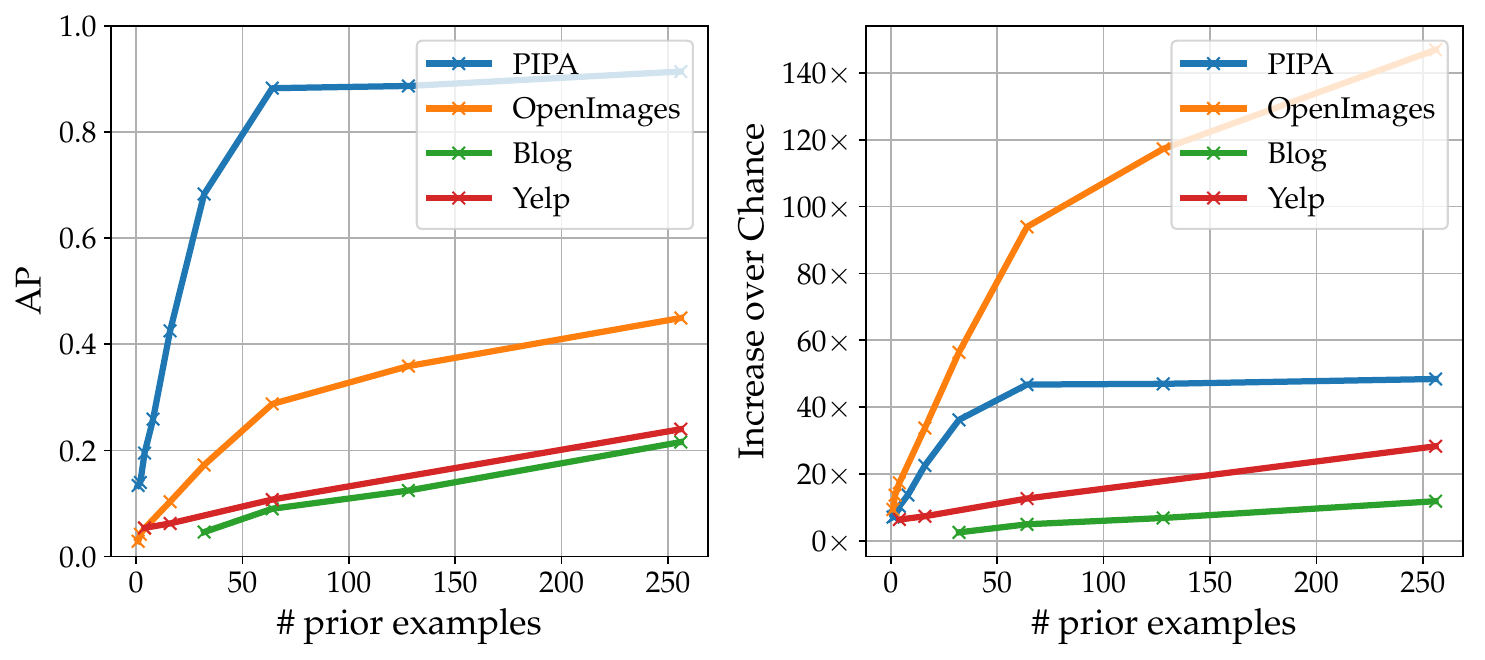}
  \caption{\textbf{Number of prior examples per user.} Evaluated on closed-world re-identification.}
  \label{fig:nexamples}
\end{figure}

We study the influence of data-limitation in deanonymization attacks in a closed-world re-identification scenario.
We previously used the entire reserve set of prior information to perform the deanonymization attacks.
We first address the influence in the amount of this prior information available per target user.
From Figure \ref{fig:nexamples}, we observe:
\begin{enumerate*}[label=(\roman*)]
  \item even a single prior example of the user leads to non-chance-level re-identification, with as much as 13.4\% AP (7 $\times$) performance on PIPA;
  \item performance of the attack increases significantly with the size of prior knowledge across all datasets e.g., 67\% increase in performance on OpenImages by using 16$\rightarrow$32 prior examples;
  \item some tasks require more prior information than others. For instance, although Blog and PIPA contain similar number of users, an adversary requires approximately 5$\times$ as many prior Blog examples to achieve 20\% AP. We attribute this to a weaker signal generated from sparse text content in Blog, as compared to dense pixel content in PIPA.
\end{enumerate*}

\begin{figure}[t]
  \centering
  \includegraphics[width=\linewidth]{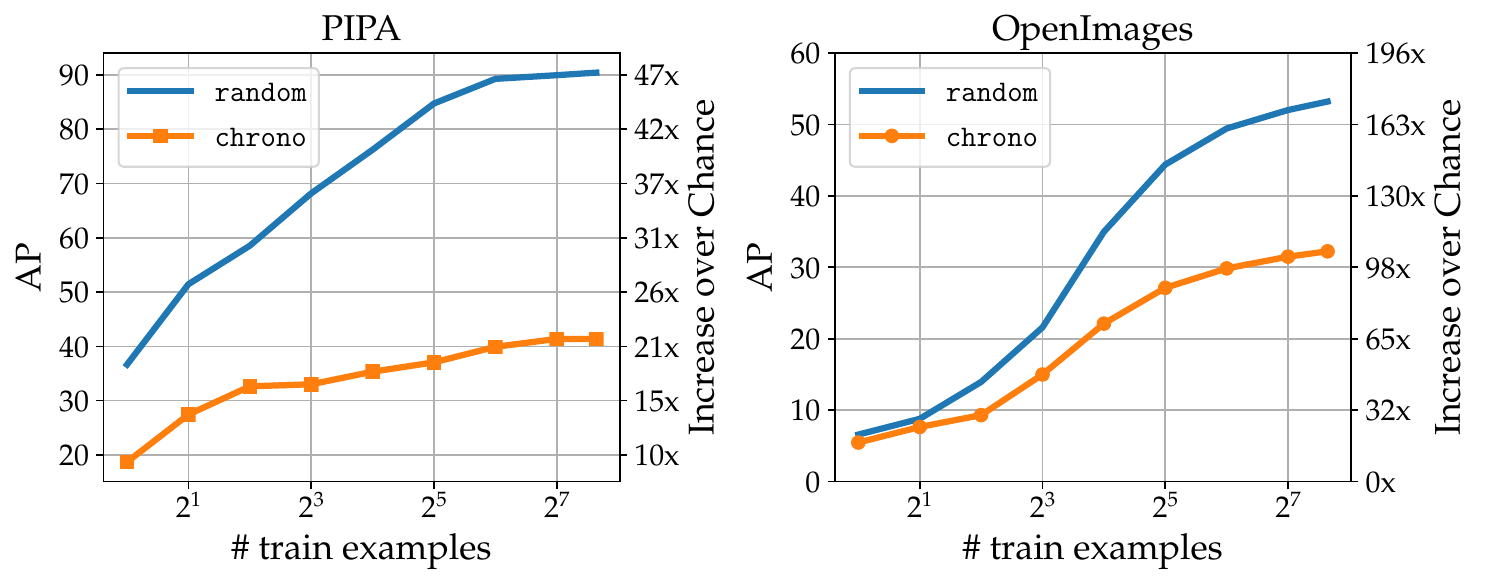}
  \includegraphics[width=\linewidth]{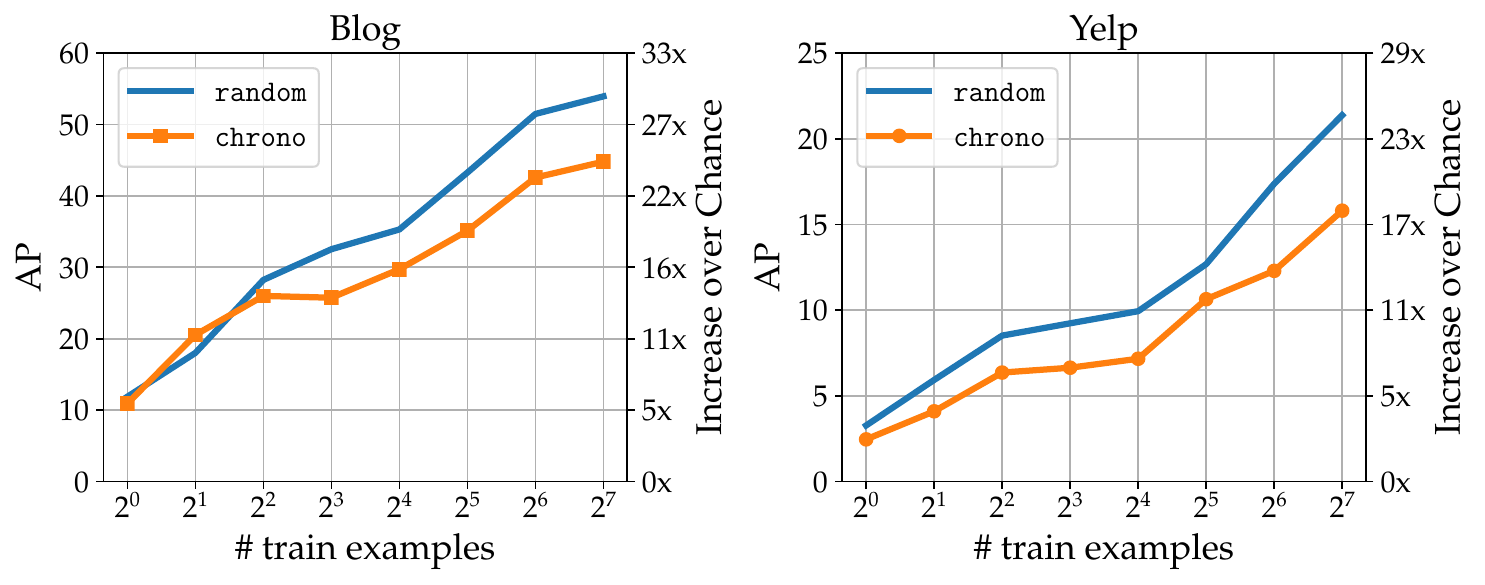}
  \caption{\textbf{Number of training examples per user.} Evaluated on closed-world re-identification.}
  \label{fig:adv_ntrain_examples}
\end{figure}

We also address the impact of size of training set ($\{\Delta \vecw_k^t : k \in \clientset_{\text{anon}}\}$) for attack models.
We train multiple re-identification MLP adversary models, each trained on a random subset of training data with increasing sizes.
In Figure \ref{fig:adv_ntrain_examples}, we observe an adversary can train reasonably effective attack models, even with extremely limited labeled data.
In particular, attack performances of 3-22$\times$ can be obtained with a single labeled example per user.
While the amount of data (either training or prior) does strongly influence the attack performance, we nonetheless find deanonymization is possible in strongly data-limited situations.

\subsubsection{Impact of Parameter Layers}
\label{sec:eval_ext_layer}
The deanonymization targets (i.e., model updates $\Delta \vecw$) comprise of parameters from multiple layers of a deep neural network.
We now analyze how the layer type and depth affect attacker performance, since they influence the type of task-specific information learnt by the model.
For instance, in CNNs, layers at various depths of the network are known to learn various concepts \cite{zeiler2014visualizing} -- lower level features (e.g., corners, edges) in the initial layers and higher level features (e.g., wheel, bird's feet) in the final layers.
For parameters updates contributed by each individual layer, we train a total of 27 attack models for CNN-based models and 3 attack models for LSTM-based models.
We were limited by storage capacity to evaluate on OpenImages as it would require $>3$TB.

\begin{figure}
  \centering
  \includegraphics[width=\linewidth]{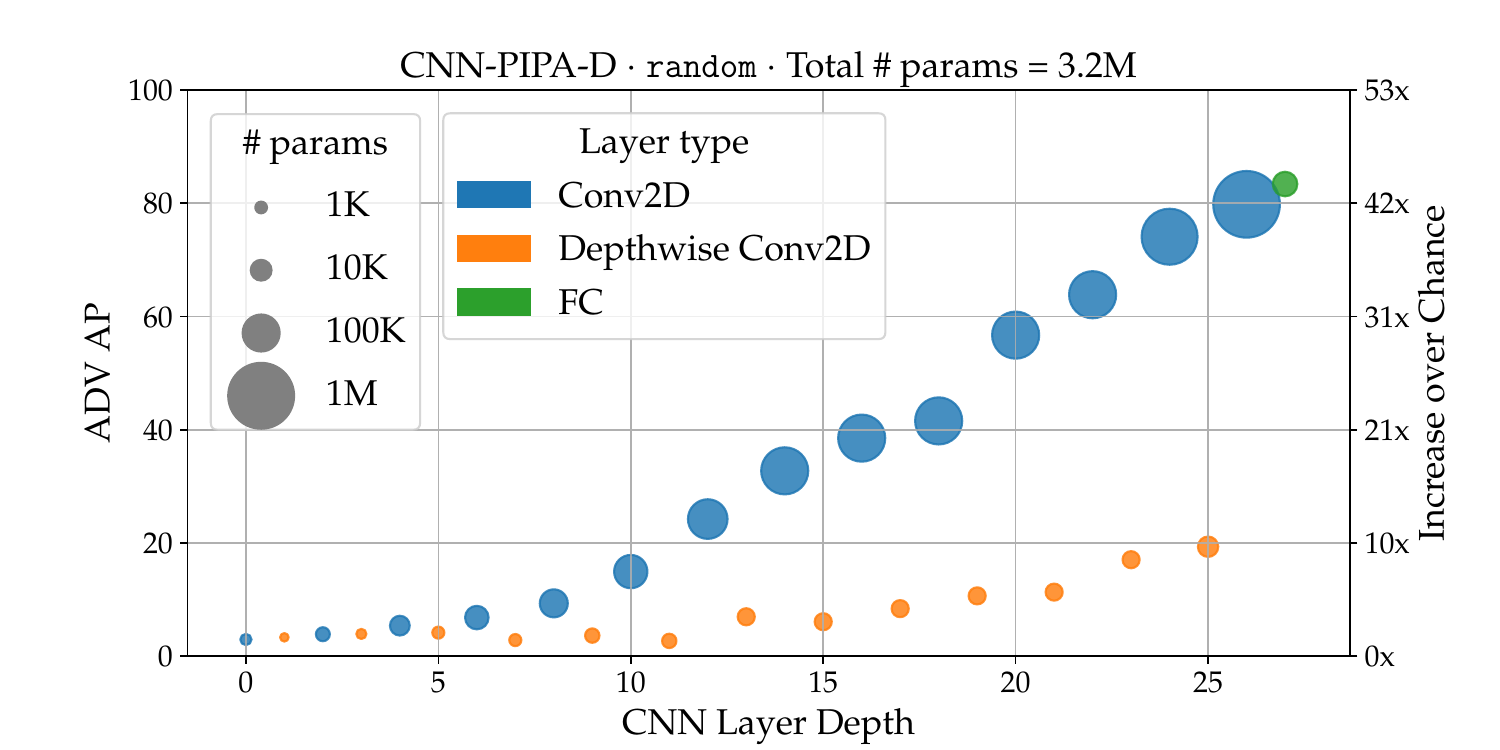}
  \caption{\textbf{Re-identification performance by depth. }
  Bubble sizes indicate the number of parameters in each layer. 
  Last two layers contains 1M and 19K parameters respectively.}
  \label{fig:ap_by_depth_cnn}
\end{figure}

\begin{table}[t]
  \scriptsize
  \begin{tabular}{@{}ccccccc@{}}
  \toprule
        &           & \multicolumn{2}{c}{NNLM-D (92K)} &  & \multicolumn{2}{c}{NNSM-D (141K)} \\ \cmidrule{3-4} \cmidrule{6-7}
  Depth & Layer type    & AP       & \# params     &  & AP       & \# params     \\ \midrule
  1     & Embedding     & 15.7 (9$\times$)     & 50K         &  & 23.5 (28$\times$)     & 50K         \\
  2     & LSTM & 46.0 (25$\times$)     & 10K         &  & 19.2 (23$\times$)     & 91K         \\
  3     & FC     & 38.8 (21$\times$)     & 32K         &  & 17.6 (21$\times$)     & 128           \\ \bottomrule
  \end{tabular}
  \caption{\textbf{Re-identification performance by depth}. For models trained on Blog and Yelp.}
  \label{tab:ap_by_depth_lstm}
\end{table}

From layer-wise performances in Figure \ref{fig:ap_by_depth_cnn} and Table \ref{tab:ap_by_depth_lstm}, we observe:
\begin{enumerate*}[label=(\roman*)]
  \item \textit{all} layers provide above-chance level information to perform re-identification attacks;
  \item in the CNN model, higher level layers contain more identifiable information with the final fully connected (FC) layer being the most informative; 
  \item in the RNN-based models, the LSTM parameters are more informative for language modeling, whereas it is the embedding layer for sentiment analysis.
\end{enumerate*}

\subsubsection{Impact of Optimization State}
\label{sec:eval_attack_round}
\begin{figure}[t]
  \includegraphics[width=\linewidth]{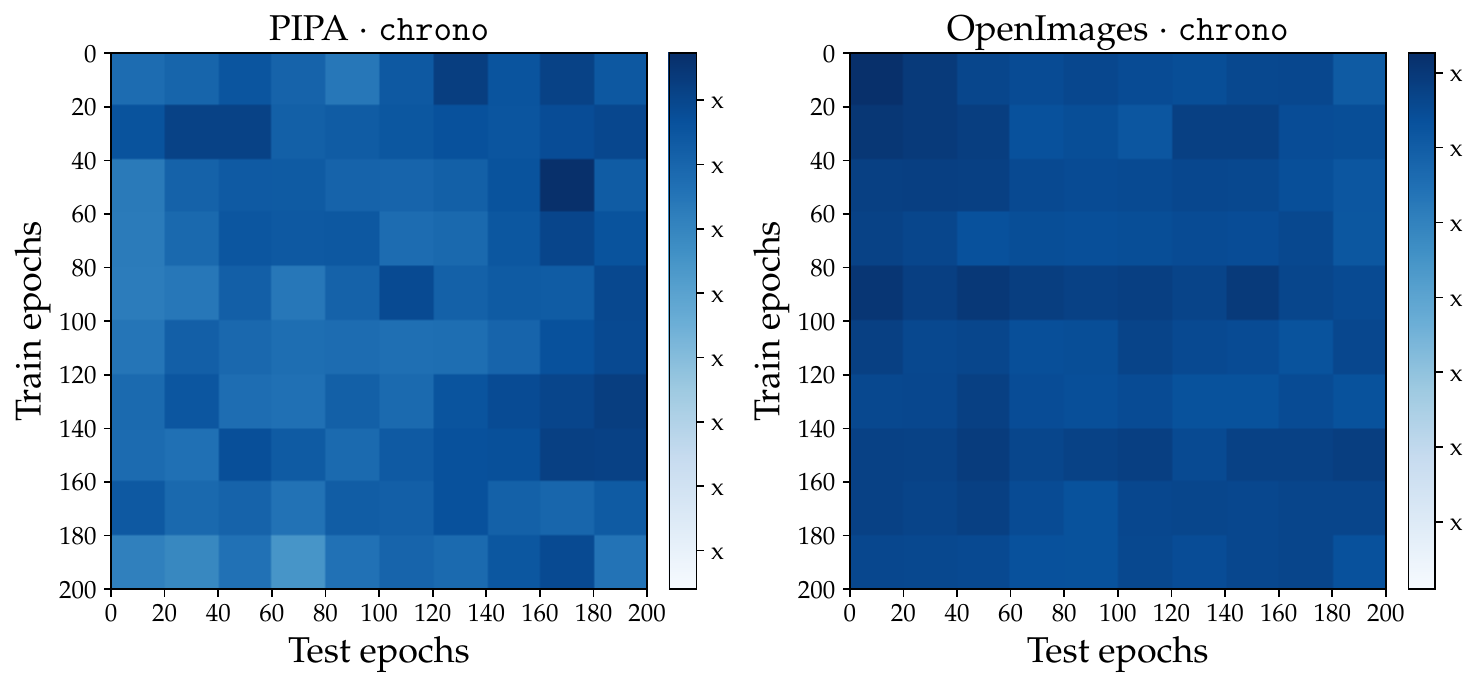}
  \caption{\textbf{Effect of the epoch $t$.} On the re-identification attack $\Delta\vecw_{\anon}^t \rightarrow u$. 
  As an example, the top-right cell denotes when the MLP was trained on $\Delta\vecw^t_u, t \in [0, 20]$ and evaluated on $\Delta\vecw^{t'}_{\anon}, t' \in [180, 200]$ }
  \label{fig:online_confmat}
\end{figure}

We now analyze the influence of training progress 
of the ML model on deanonymization attacks.
We group the parameter updates (separately for train and test attack sets), based on the epoch ranges during which they were generated.
We split parameter updates collected during training of $f_{\vecw}$ over 200 epochs into 10 ranges, each with 20 epochs.
We train and evaluate the MLP re-identification attack model over all $10\times 10$ train-eval pairs.
From Figure \ref{fig:online_confmat}, we observe that the training progress at which the update was generated has little influence on the performance indicating an adversary can re-identify users at any stage of training.

\subsubsection{Reasoning About Effectiveness of Attacks}
\label{sec:eval_reason_biases}

\begin{figure}
    \centering
  \includegraphics[width=0.8\linewidth]{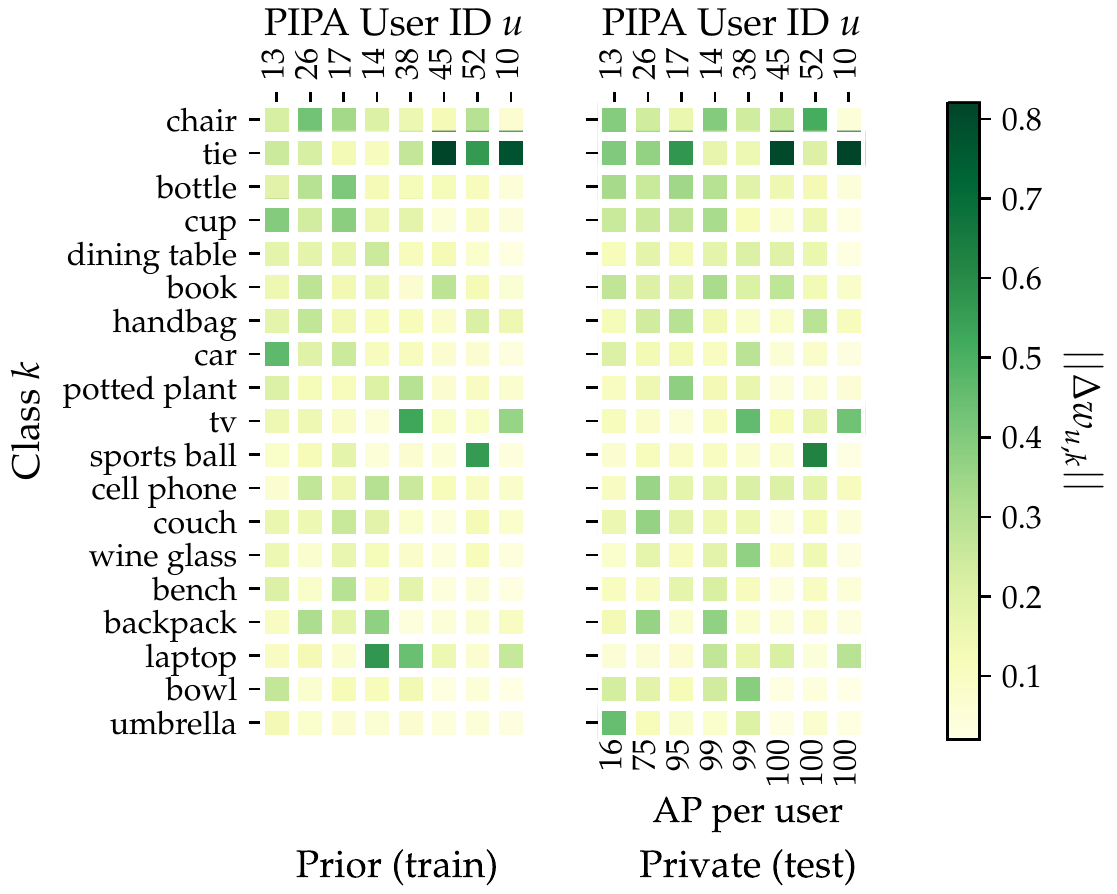}
  \caption{\textbf{User bias visualized on parameter updates.} 
  }
  \label{fig:fcnorms_pipa}
\end{figure}

In Section \ref{sec:threat_bias} (Fig. \ref{fig:dataset_distances}), we observed that users display a bias resulting in lower variations in data they capture.
Consequently, we conjectured that the resulting bias is consistently encoded in the parameter updates, even when they are computed on different (prior and private) sets of users' data.
To validate, we take a closer look at the parameter updates $\Delta \vecw_u^{\text{prior}}, \Delta \vecw_u^{\text{private}} \in \real^{D \times K}$ in the FC layer of eight users in the PIPA FL setup, where $K$ (=19) is the number of classes and $D$(=1024) represents weights per class.
In Figure \ref{fig:fcnorms_pipa}, we illustrate bias per user (columns) in the parameter delta space by computing the $L_2$-norm of each of the $K$ class weight vectors (column-dimension in $\Delta \vecw_u$).
We observe:
\begin{enumerate*}[label=(\roman*)]
  \item for users who can be re-identified highly accurately (e.g., $u$=10), we find that the user is more biased towards images containing `tie', `tv', and `laptop'. Furthermore, this bias is consistent in both the user's prior and private update signals; and 
  \item surprisingly, even when biases are not entirely consistent (e.g., $u$=17), we find attacks to be reasonable effective (AP=95); and 
  \item for users who cannot be re-identified easily (e.g., $u$=13), the biases are inconsistent between the prior (biased towards cars and cups) and private (biased towards chairs, ties, and umbrellas) update signals.
\end{enumerate*}
We find our conjecture that the user bias signal translates to the parameter delta space, holds reasonably well, leading to highly effective deanonymization attacks we saw in the previous sections.

\section{Countermeasures}
\label{sec:mitigation}
In the previous section, we evaluated our threat models across a variety of challenging scenarios and consistently observed deanonymization risks.
In this section, we present mitigation strategies to counter these attacks.

We attributed (\S \ref{sec:eval_reason_biases}) the effectiveness of the attacks to user bias, which is a powerful statistical signal in both the limited set of adversary's prior data and the users' private data. 
The focus of our mitigation strategies is to perturb the data bias on the anonymous device, to provide a false signal to the adversary.
We spell out our requirements for the defense as:
\begin{enumerate*}[label=(\alph*)]
  \item maximally retain utility (performance of $f_{\vecw}$);
  \item involve low computation overhead;
  \item not rely on a trusted-third party; and
  \item allow users to selectively employ the strategy to various extents depending on personal preferences.
\end{enumerate*}

\subsection{Methods}
Based on the requirements, we propose data-centric mitigation strategies: devices adversarially bias their data distribution on devices, rather than directly perturb model parameters. 
More specifically, users mix their original data $\data_u$ with certain ``background'' data $\bkg$ to ``blend into the crowd'', thereby rendering the parameters less user-specific. 
Here, the mixing takes place prior to participation in FL.

\begin{table}[]
  \scriptsize
  \begin{tabular}{@{}lllll@{}}
  \toprule
  $\data$    & Source ($\data$) & $\bkg$ & Source ($\bkg$) & $|\bkg|$ \\ \midrule
  PIPA \cite{piper}       & Flickr     & OpenImages         & Flickr        & 59K  \\
  OpenImages \cite{openimages} & Flickr     & OpenImages         & Flickr        & 490K \\
  Blog \cite{schler2006effects}      & Blogger    & WikiReading \cite{hewlett2016wikireading}       & Wikipedia     & 3M   \\
  Yelp \cite{challenge2013yelp}      & Yelp       & Amazon Reviews \cite{he2016ups}     & Amazon        & 1.7M \\ \bottomrule
  \end{tabular}
  \caption{\textbf{Background datasets and sources.} Used to mitigate deanonymization attacks.}
  \label{tab:mitigation_datasets}
\end{table}

\myparagraph{Collecting $\bkg$}
The background dataset $\bkg$ can be any large (labeled) set of training examples for the same federated learning task (e.g. user-annotated dataset, scraped data from the Internet, a trusted open-source dataset).
The background datasets used in our experiments, their sources and sizes are listed in Table \ref{tab:mitigation_datasets}.
We only select a random subset of the original background datasets (s.t. $|\bkg| \gg |\data_u|$) in each case, for experiments to complete within a feasible amount of time.
The preprocessing of $\bkg$ and $\data$ are identical.

Now, we present three countermeasures which alter the characteristic data-distribution of the users.

\myparagraph{Data Replacement (\texttt{bkg-repl})}
Each user replaces a fraction $\alpha \in [0, 1]$ of his/her data $\data_u$ with ones from $\bkg$.
At $\alpha = 0$, no mitigation strategy takes place; at $\alpha = 1$, every user has identical data composition.
However, the strategy skews FL to learn from a noisy background data distribution displaying different statistics, instead of learning from interesting user data on which evaluation metrics need to be maximized.

\myparagraph{Data Augmentation (\texttt{rand-aug})}
Instead of replacing, the user \emph{augments} random data (since more data helps \cite{sun2017revisiting,halevy2009unreasonable}) from $\bkg$:
\begin{equation}
	\hat{\data_u} \leftarrow \data_u \cup \{(\vecx_i, \vecy_i) \sim \bkg\}_{i=1}^{\alpha \cdot |\data_u|},
\end{equation}
where $\alpha \ge 0$ determines the size of augmentation.
As $\alpha \to \infty$, devices' empirical data distributions converge to $\bkg$, making them indistinguishable from each other.

\myparagraph{Mode-specific Data Augmentation (\texttt{mm-aug})}
So far, the users' strategies were to mix their data with background data from a single source $\bkg$.
We now consider the strategy where each device mixes data from \textit{different} topics i.e., modes of the data distribution.
For instance, Alice adversarially adds sports content to her data to mask her interest in automobiles before participating in FL.
We perform this by first clustering $\bkg$ into $M$ clusters \linebreak $\bigcup_{m=1}^{M} \bkg_m$. We use the k-means clustering over the ImageNet pretrained Mobilenet features.
Each user $u$ picks a cluster $m$ at random,
and augments its data with ones from the cluster:
\begin{equation}
	\hat{\data_u} \leftarrow \data_u \cup \{(\vecx_i, \vecy_i) \sim \bkg_m\}_{i=1}^{\alpha\cdot|\data_u|}
\end{equation}
where $\alpha\geq 0$ controls the degree of mix. 
We use $M$=100 for PIPA, $M$=500 for OpenImages, $M$=300 for Blog and Yelp.

We additionally consider two perturbation-based baselines to our data-augmentation strategies.

\myparagraph{DP-FederatedAveraging (\texttt{dp-fedavg})}
We implement a differentially private variant \cite{mcmahan2017learning} of the Federated Averaging algorithm.
They key idea is to provide $(\varepsilon, \delta)$ participant-level differential privacy guarantees by bounding the contribution (the parameter update) provided by each participant.
In practise, the contributions are bounded by clipping the parameter updates and further adding random noise.
In our experiments, we fix the clipping value to 50 and vary magnitude of gaussian noise added during training.

\myparagraph{Random Perturbations (\texttt{noise})}
Although \texttt{dp-fedavg} has shown success in large-scale scenarios (with thousands of users), we found difficulty achieving reasonable results in our setup.
Hence, we consider a relaxed version of introducing perturbations, where the user introduces zero-centered Gaussian noise to model updates before leaving the device.

\subsection{Evaluation}
We evaluate the proposed mitigation strategies by measuring the adversary's performance against our countermeasures. 
We analyze the effectiveness of the defense against the strongest adversary: closed-world re-identification attack on \texttt{random} prior (\S \ref{sec:eval_attack_priordist}, Table \ref{tab:eval_identification_closed}).

\begin{figure}
  \includegraphics[width=\linewidth]{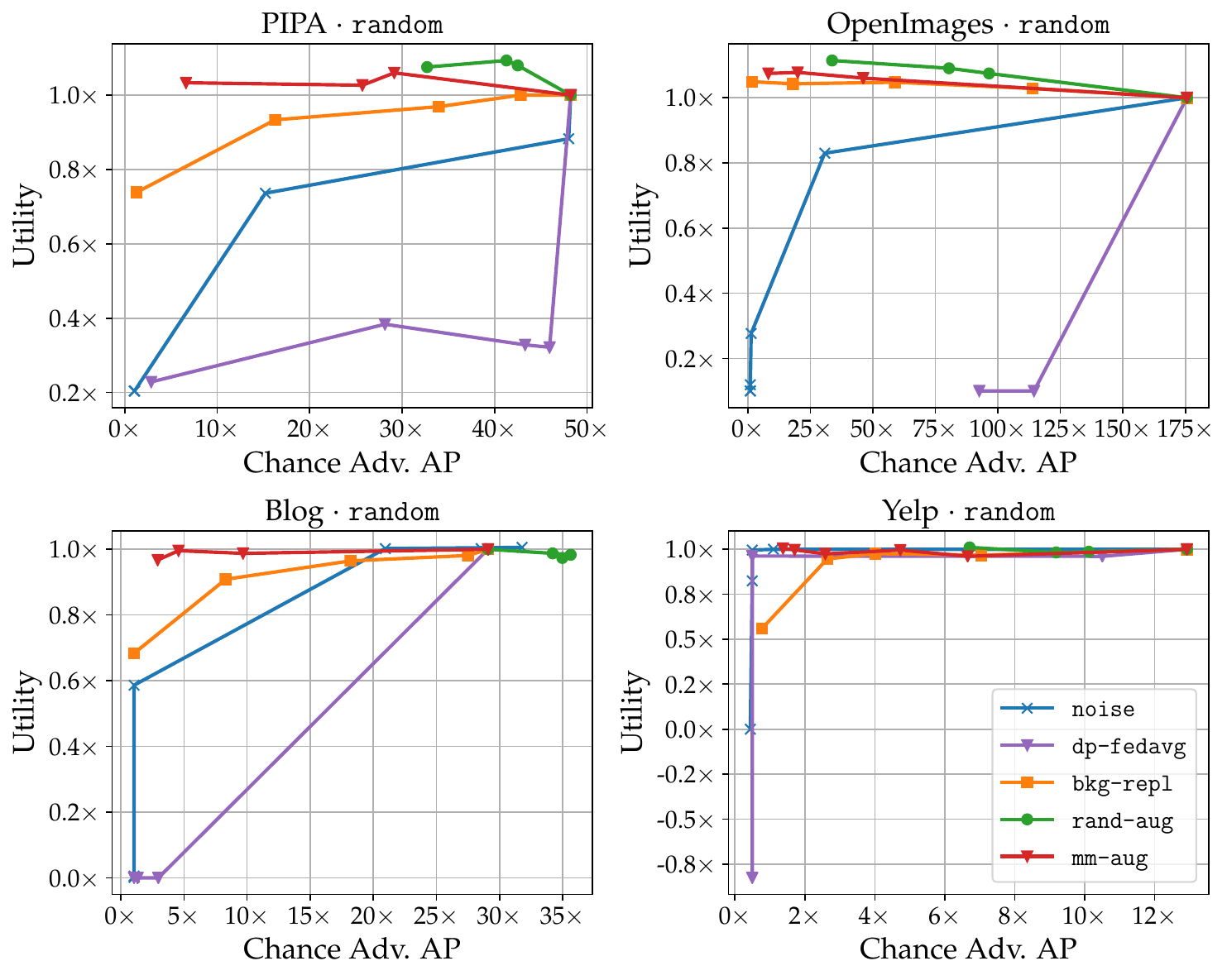}
  \caption{\textbf{Mitigation strategies evaluation.} Re-identification AP obtained by varying $\alpha$ and $\sigma^2$ in closed-world scenario. Top-left is the ideal region. Higher $\alpha$ and $\sigma^2$ values pushes operating points towards the left (i.e., lower deanonymization performance).}
  \label{fig:mitigation_plots_comb}
\end{figure}

We evaluate the strategies in terms of trade-off between privacy (reduction in adversary's performance) and utility (decentralized learning performance).
As in \S \ref{sec:eval_attack_priordist}, we measure the adversary's performance as increase over chance-level AP.
We measure utility by performance scores normalized to have utility=$1.0$ when no mitigation takes place.

The mitigation strategies are evaluated on a curve by varying hyperparameters.
For \texttt{bkg-repl}, we use $\alpha \in$ \{0.0, 0.25, 0.50, 0.75, 1.0\}.
For \texttt{rand-aug} and \texttt{mm-aug}, we use $\alpha \in$ \{0.0, 0.5, 1.0, 2.0\}.
For \texttt{dp-fedavg}, we fix the clip value to 50 and vary the noise multiplier in the range $[10^{-5}, 10^{-1}]$.
For \texttt{noise}, we consider Gaussian noise with $\mu = 0$ and $\sigma^2 \in \{10^{-2}$, $10^{-1}$, $10^{0}$, $10^{1}$, $10^{2}\}$.

We present evaluation for our strategies in Figure \ref{fig:mitigation_plots_comb}.
Better mitigation strategies have curves towards the top-left corners in each plot (high privacy, high utility).
We observe:
\begin{enumerate*}[label=(\roman*)]
  \item the perturbation-based baselines (\texttt{noise} and \texttt{dp-fedavg}) in most cases severely decreases utility at a small gain in privacy; 
  \item replacing data with background samples (\texttt{bkg-repl}) is a good alternative strategy: we have both higher privacy and utility than perturbation methods. 
  However, due to a domain-shift between $\bkg$ and $\data$, utility is often impacted.
  This can be observed in PIPA, Blog and Yelp datasets, where it achieves $<0.75\times$ utility since the user data is no longer used;
  \item the augmentation-based strategies \texttt{rand-aug} and \texttt{mm-aug} outperforms \texttt{noise} and \texttt{bkg-repl} in terms of utility and privacy; 
  \item for the \texttt{mm-aug} strategy, already at $\alpha=0.5$, we observe a good combination of privacy and utility (75\% decrease in adversary's AP in OpenImages, compared to 45\% for \texttt{rand-aug} and 67\% for \texttt{bkg-replace}).
\end{enumerate*}

We find the strategy \texttt{mm-aug} offer the most effective and practical operating points, requiring the user to perform minimal augmentation to achieve reasonable privacy. 
We remark that the utility for \texttt{mm-aug} can be more than $1.0$ even at higher privacy level, as can be seen in PIPA and OpenImages.
This is due to the effect of additional data \cite{halevy2009unreasonable,sun2017revisiting}.
This increased privacy and utility comes at the cost of preparing a labeled dataset and increased training time (training set becomes $(1 + \alpha)\times$ bulky).
However, this overhead will be less costly with increasingly powerful devices and energy-efficient ML models for mobile devices \cite{howard2017mobilenets,sandler2018inverted}.

\section{Conclusion}
In this paper, we were motivated to understand privacy threats in Federated Learning, which is designed towards large-scale learning on user data on personal devices.
We questioned whether devices can truly participate anonymously without compromising the identity of individuals.
Our results indicate that the devices can be effectively deanonymized using the transmitted model parameter updates and a reasonable amount of prior data.
We found this to be possible due to the inherent user bias in captured data acting as a fingerprint that is consistent across different sets of data captured by the user.
To mitigate such attacks, we proposed calibrated domain-specific data augmentation, which shows strong results in preventing deanonymization with minimal impact to utility.

\bibliographystyle{plain}
\bibliography{main}

\end{document}